\documentclass[twocolumn,english,aps,showpacs,nofootinbib]{revtex4}
\usepackage[T1]{fontenc}
\usepackage[latin9]{inputenc}
\usepackage{amsmath}
\usepackage{amssymb}
\usepackage{esint}
\usepackage{graphicx} 
\usepackage{verbatim} 
\makeatletter
\@ifundefined{textcolor}{}
{%
 \definecolor{BLACK}{gray}{0}
 \definecolor{WHITE}{gray}{1}
 \definecolor{RED}{rgb}{1,0,0}
 \definecolor{GREEN}{rgb}{0,1,0}
 \definecolor{BLUE}{rgb}{0,0,1}
 \definecolor{CYAN}{cmyk}{1,0,0,0}
 \definecolor{MAGENTA}{cmyk}{0,1,0,0}
 \definecolor{YELLOW}{cmyk}{0,0,1,0}
 }

\usepackage{nicefrac}\usepackage{dsfont}
\usepackage[normalem]{ulem}\@ifundefined{definecolor}
 {\usepackage{color}}{}

\DeclareMathOperator\erf{erf}
\usepackage{enumerate}

\makeatother

\usepackage{babel}
\begin{document}

\title{On-site residence time in a driven diffusive system: violation and recovery of mean-field}
\author{J. Messelink}
\author{R. Rens}
\author{M. Vahabi}
\author{F. C. MacKintosh}
\author{A. Sharma}
\affiliation{Department of Physics and Astronomy, VU University, Amsterdam, The Netherlands}


\date{\today}
\begin{abstract}
We investigate simple one-dimensional driven diffusive systems with open boundaries. We are interested in the average on-site residence time defined as the time a particle spends on a given site before moving on to the next site. Using mean-field theory, we obtain an analytical expression for the on-site residence times. By comparing the analytic predictions with numerics, we demonstrate that the mean-field significantly underestimates the residence time due to the neglect of time correlations in the local density of particles. The temporal correlations are particularly long-lived near the average shock position, where the density changes abruptly from low to high. By using Domain wall theory (DWT), we obtain highly accurate estimates of the residence time for different boundary conditions. We apply our analytical approach to residence times in a totally asymmetric exclusion process (TASEP), TASEP coupled to Langmuir kinetics (TASEP + LK), and TASEP coupled to mutually interactive LK (TASEP + MILK). The high accuracy of our predictions is verified by comparing these with detailed Monte Carlo simulations.
\end{abstract}

\maketitle


\section{Introduction}
Driven diffusive systems have been intensively studied in the last three decades. These systems, being inherently nonequilibrium, exhibit very rich dynamical as well as steady state behavior~\cite{krug1991boundary, derrida1993exact, evans1995spontaneous,domb1995statistical, evans1998phase,o1998jamming,schmittmann1998driven,helbing2001traffic,parmeggiani2003phase, parmeggiani2004totally,kwak2004driven}. A considerable theoretical effort has been put into modeling such systems. One particular model that has similar appeal to the Ising model for equilibrium physics is the totally asymmetric simple exclusion process (TASEP) in 1D~\cite{krug1991boundary, derrida1993exact, evans1995spontaneous, evans1998phase, kafri2002criterion}. As shown in Fig.~\ref{lkmodel}, in this model a single species of particles performs unidirectional hopping on a 1D lattice. Due to hard-core repulsions an empty site can be occupied by a single particle only. 

TASEP and its variants have been used to model intracellular transport~\cite{appert2015intracellular}, motion of Ribonucleic polymerase (RNA polymerase) on the DNA (Deoxyribonucleic acid) during transcription and of Ribosomes on the RNA during translation~\cite{shaw2003totally,ciandrini2010role,basu2007traffic,reuveni2011genome,heinrich1980mathematical,chowdhury2005physics}. Advances in experimental methods to track macromolecules moving on 1D tracks inside a cell~\cite{kural2007tracking,churchman2005single,yildiz2005fluorescence,courty2006tracking,harada2001direct,skinner2004promoter,andrecka2008single,adelman2002single} have enabled a potentially direct link between theoretical work on the TASEP model and experiments.

In this study we focus on the on-site residence time of system particles, defined as the average amount of time a particle spends at a particular site. Our study is directly relevant to the residence time of motor proteins moving on biofilaments inside a cell, as one can experimentally measure the time that an attached motor spends on the filament by labeling motor proteins with fluorophores. The study of on-site residence times is closely related to the studies on dynamics of a tagged particle in an exclusion process~\cite{kipnis1986central,ferrari1996poissonian,imamura2007dynamics}. In these studies the focus is on the distribution of a tagged particle's position performing exclusion process on a 1D infinite lattice. The position of the tagged particle in time is directly related to the on-site residence time of the tagged particle. In this article we especially focus on the on-site residence times of a TASEP system that exhibits a first-order-like phase transition. This transition, which occurs for a certain set of boundary conditions, comprises of two phases, low and high density (LD and HD), which coexist on a lattice separated by a shock-interface~\cite{krug1991boundary, evans1995spontaneous, evans1998phase, kafri2002criterion}. The interface can be delocalized over the entire lattice \cite{kolomeisky1998phase,santen2002asymmetric,rakos2003hysteresis}. However, in a modified version of TASEP, in which the number of particles is not conserved on the lattice due to adsorption (desorption) from (to) an infinite reservoir, the shock interface is localized~\cite{parmeggiani2003phase, parmeggiani2004totally}. This model is known as the TASEP coupled to Langmuir kinetics (LK) which we refer to as the TASEP + LK model (see Fig.~\ref{lkmodel}). A further modification in which TASEP is coupled to the mutually interactive Langmuir kinetics (TASEP + MILK) can lead to even stronger localization than the TASEP + LK model~\cite{vuijk2015driven,ebbinghaus2011particle}. Despite the interaction with the bulk, the 1D system exhibits phase transition behavior under the mesoscopic scaling of kinetic rates (attachment/detachment rates from the lattice). By making the cumulative kinetic rates independent of the lattice size, boundaries can compete with the bulk even in the thermodynamic limit~\cite{parmeggiani2003phase, parmeggiani2004totally}. This mesoscopic scaling of the kinetic rates is potentially applicable to molecular motors performing directed motion along 1D molecular tracks inside cells. Typically, kinetic rates are such that the motors move on a significant fraction of the track before undergoing detachment~\cite{howard2001mechanics}. This allows for the bulk dynamics to compete with the boundary, potentially giving rise to unusual nonequilibrium stationary states.

The on-site residence time of a particle performing TASEP, when summed up over the entire lattice of a given size yields the total residence time of a typical particle between the injection event at one boundary to the extraction event at the other boundary. By relating the residence time to the size of the lattice, one can obtain transport coefficients analogous to mobility and diffusion constant. In the presence of Langmuir kinetics, the total residence time will not be equal to the sum of the on-site residence times as the possibility of a particle detaching before reaching the other end of the lattice also needs to be taken into account. In this study we focus on the on-site residence time of TASEP as well as TASEP+LK systems. We first present a mean-field description of the on-site residence time. Mean-field theory predicts that the residence time depends only on the steady-state density of particles. However, we show that the mean-field estimate significantly underestimates the residence time of systems in LD-HD phase (the two-phase coexistence state mentioned previously) due to neglect of time correlations in the density. The time correlations, irrelevant in calculation of the steady-state density profile, become important in calculation of the residence time. We show that the time correlations are especially significant near the average location of the shock interface. We present a non mean-field theory of calculating the on-site residence time. As an input to our theory, we need the spatial distribution of the shock interface in the steady state which we obtain using the domain wall theory~\cite{kolomeisky1998phase,santen2002asymmetric}. We present analytical estimates of the residence time and demonstrate, comparing with numerics, the high accuracy of our theory.

The article is organized as follows. In Sec.~\ref{model}, we present the model considered in this study. We briefly summarize the findings of previous studies on TASEP as well as TASEP + LK which are relevant to our study. In Sec.~\ref{mean-field}, we present mean-field theory for calculation of the on-site residence time. We neglect any time correlations and demonstrate that the on-site residence time is only dependent on the steady state density profile.
In Sec.~\ref{ourtheory}, we present non mean-field theory for calculating residence time. The derivation of the non mean-field theory makes use of two assumptions on the movement of the shock interface in LD-HD phase. These assumptions are supported by investigating the flag model for the movement of the shock in Sec.~\ref{DWT} and extending this model to TASEP+LK. In Sec.~\ref{results} we present our theoretical estimates for the residence time and compare them with numerics. We apply our theoretical anlysis to TASEP (with and without LK). 
In Sec.~\ref{MILKsection}, we present a general approach to residence time by introducing a probe particle and apply our findings to the TASEP + MILK case. In Sec.~\ref{limitations}, we present a transition from our predicition of residence times to the mean-field theory prediction by breaking one of the assumptions on the movement of the shock interface.  
Finally in Sec.~\ref{conclusions}, we summarize our findings and present a brief outlook.


\section{The model} \label{model}
The TASEP model is described as follows. The system comprises of a 1D lattice of length $L$ and $N$ sites. Particles are injected into the lattice from the leftmost site with a rate of $\alpha$ and detach from the rightmost site with a rate of $\beta$. The totally asymmetric motion of a particle corresponds to its motion exclusively to the right. The conventional parameter $p$ for the hopping rate on the lattice is set to one here. Every particle is subjected to hard-core repulsion. Each site is thus either unoccupied, or occupied by a single particle. 

\begin{figure}[h]
\includegraphics[width = \columnwidth]{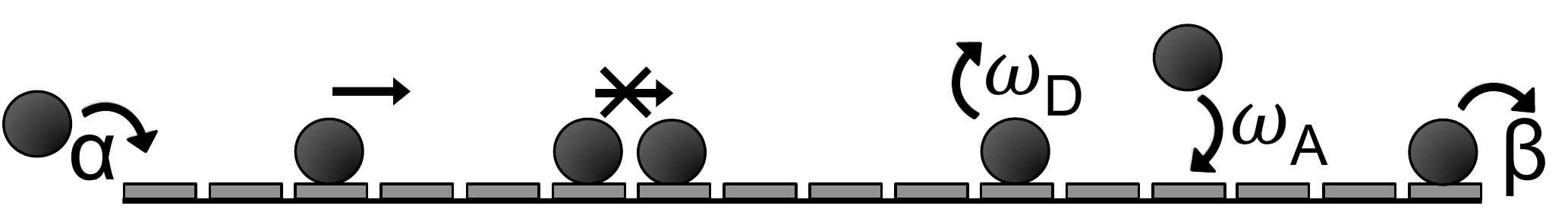}
\caption{Schematic of the TASEP model with Langmuir kinetics. Particles are injected at the first site with rate $\alpha$ and extracted at the last site with rate $\beta$. The motion of particles is totally asymmetric as they move exclusively to the right with a unitary rate given that the site is empty. Hopping over an occupied site is not allowed. Pure TASEP is comprised of injection, extraction, and asymmetric hopping of particles interacting via hard core repulsion. In the TASEP + LK model, Langmuir kinetics are additionally included by allowing particles to detach from an occupied site with rate $\omega_D$, and attach to an unoccupied site with rate $\omega_A$.}
\label{lkmodel}
\end{figure}

The TASEP model coupled to Langmuir kinetics is an extension of this model, in which each particle has a probability of $\omega_D$ of detaching during a single timestep, and each empty site has a probability of $\omega_A$ of a particle attaching to it from the bulk reservoir during a single timestep (See Fig.~\ref{lkmodel}). We assume that the bulk reservoir has an infinite capacity to act as both source and sink of particles. We see that the original TASEP model is recovered if we set $\omega_A = \omega_D = 0$.

The sites are indexed from $i=0$ at the injecting boundary to $i = N$ at the extracting boundary. The total length of the lattice is $L$ and thus each site has a spatial extent of $\epsilon = L/N$. Under the assumption that in steady-state spatial correlations can be ignored, one can derive the following mean-field equation for the average density $\rho(x)$ of particles~\cite{parmeggiani2003phase, parmeggiani2004totally}:

\begin{equation}\label{diffLK}
\frac{\epsilon}{2} \partial_x^2 \rho + (2\rho-1)\partial_x \rho  + \Omega_A (1-\rho) - \Omega_D \rho = 0.
\end{equation}
In the equation above, the continuous spatial parameter $x$ runs from $0$ to $L$ and corresponds to the location of site $i$ through $x = i\epsilon$. The parameters $\Omega_A$ and $\Omega_D$ are referred to as the total attachment and detachment rates, respectively and are defined as $\Omega_A = \omega_A N$, and $\Omega_D = \omega_D N$. Mesoscopic scaling of the total kinetic rates corresponds to fixing $\Omega_A$ and $\Omega_D$ for a given $L$. Such scaling ensures that $N$ can be varied without modifying the total kinetic rates from the lattice. It is only under this scaling that one obtains boundary driven phase transition~\cite{parmeggiani2003phase} by letting $N\rightarrow \infty$ such that $\epsilon \rightarrow 0$, reducing eq.~\eqref{diffLK} to a first-order differential equation. In the thermodynamic limit of $N\rightarrow \infty$, the first-order differential equation is overdetermined as there are two boundary conditions given by $\rho (0) = \alpha$ and $\rho (L) = 1-\beta$. Setting $\Omega_A = \Omega_D = \Omega$, we have three solutions to the resulting differential equation:

\begin{align}
&\rho_{\mathrm{LD}} = \alpha + \Omega x  \label{soln1}\\
&\rho_{\mathrm{HD}} = 1-\beta-\Omega L + \Omega x \label{soln2} \\
&\rho_{\mathrm{MC}} = \frac{1}{2}. \label{soln3}
\end{align}
The solution $\rho_{\mathrm{LD}}$ is referred to as the low density (LD) phase. The solution $\rho_{\mathrm{HD}}$, as the high density (HD) phase, and $\rho_{MC}$ as the maximum current (MC) phase. In this study we focus on shock-type profiles, which one obtaines when $\Omega < 1-\beta-\alpha$ and $\alpha, \beta < \frac{1}{2}$. An example of a shock-type density profile is shown in figure \ref{timecorrs}a and \ref{timecorrs}b. The shock-type profile is a two-phase coexistence scenario in which the density makes an abrupt transition between $\rho_{\mathrm{LD}}$ and $\rho_{\mathrm{HD}}$. The mean location of the shock is determined by finding the point on the lattice where the current $\rho(x)(1-\rho(x))$ of the low density phase matches that of the high density phase~\cite{parmeggiani2003phase}. 
\\
The original TASEP model, described in Ref.~\cite{spitzer1970interaction}, is recovered by setting $\Omega = 0$. An interesting modification to the original TASEP + LK model has been recently proposed in Ref.~\cite{vuijk2015driven}. The authors considered modified LK such that the local attachment and detachment rates of a particle depend on the the state of the neighboring sites. This model is referred to as the TASEP + MILK model and exhibits qualitatively different phase behavior from the TASEP + LK model. However, one also obtains two-phase coexistence, as in TASEP + LK. We defer the discussion of TASEP + MILK model to section~\ref{MILKsection}.

The two phase-coexistence in TASEP + LK is strictly present only in the thermodynamic limit of $N\rightarrow \infty$ along with the mesoscopic scaling of the kinetic rates. In pure TASEP, however, shock interface can exist for a finite $N$ for appropriately chosen boundary conditions. For any finite $N$, the shock interface is delocalized around its mean-location~\cite{evans2003shock}. Such delocalization, in steady state, results in a smoothing of density profile from low density to high density phase. For finite $N$, it is interesting to pose the following question: if a particle is injected from the leftmost site of the lattice, how long does this particle typically stay on the lattice? And what length of time will it typically spend on each site? In particular, would the dynamics of a delocalized shock interface have important implications for the residence times? Or can one ignore the dynamics entirely and obtain the residence times in a mean-field fashion? To address these questions, we first consider the mean-field approach below.

\section{Mean-field approach to residence time} \label{mean-field}
The residence time on a lattice with $N$ sites, as defined in Ref.~\cite{cirillo2014residence} is the average time a particle entering the lattice spends on the lattice before detaching from the lattice. In this article we additionally make use of the \textit{on-site residence time}: the average length of time a particle attaching on a particular site spends on that site before making a hop or detaching. For the TASEP+LK model, we only consider the residence times of particles attaching at the $\alpha$ boundary.

We first consider a unidirectional random walker on a one-dimensional lattice, located on a site with discrete index $i$. This index runs from $i=0$ to $i=N$. At each timestep it has a probability of $p_{i+1}$ to detach from site $i$ (which could either be due to hopping to the next site, or due to a detachment event), and a probability of $1-p_{i+1}$ to stay at site $i$. One can calculate the total time that a particle spends on site $i$ before detaching from that site as follows: 
\begin{equation} \label{summation}
\sum_{n=1}^{\infty} n (1-p_{i+1})^{n-1} (p_{i+1}) = \frac{1}{p_{i+1}}.
\end{equation}
In this summation, each waiting time $n$ is weighted by the probability of the particle to detach after exactly $n$ timesteps.\\

Eq.~\eqref{summation} assumes that the detachment probability is a Poisson distribution. An explicit expression for the average on-site residence time $r_i$ of a particle on site $i$, that takes the system dynamics into account can be written as
\begin{widetext}
\begin{align}
r_i &= \bigg\langle (1-\omega_D)(1-N_{i+1}(t))+\omega_D \begin{aligned}[t]
&+ 2 N_{i+1}(t)(1-\omega_D)[(1-N_{i+1}(t+1))(1-\omega_D)+\omega_D] \\ 
&+ 3 N_{i+1}(t)N_{i+1}(t+1)(1-\omega_D)^2 [(1-N_{i+1}(t+2))(1-\omega_D)+\omega_D]+...\bigg\rangle \nonumber \\
\end{aligned}\\
&= \left\langle 1-N_{i+1}(t)(1-\omega_D) + \sum_{j=1}^{\infty} (j+1)(1-N_{i+1} (t+j)(1-\omega_D))(1-\omega_D)^j \prod_{k=0}^{j-1} N_{i+1} (t+k)\right\rangle. \label{explicit}
\end{align}
\end{widetext}
Here, $N_i(t)$ denotes the occupation of the site $i$ at time $t$ and can be either 0 or 1. The symbol $t$, used throughout the article, denotes discrete time and advances in units of 1. The average is to be understood as an average over several realizations of the same system in steady state. In the rest of this article we refer to the quantity between the brackets in eq. \eqref{explicit} as $\tau_i (t)$, i.e. we write
\begin{equation}
r_i = \langle \tau_i (t) \rangle.
\end{equation}

In order to obtain a closed form expression for the on-site residence time, the time correlations of the form $\langle \prod\limits_{k=0}^{n}N_j (t+k) \rangle$ are needed for all $n \in \mathbb{N}$. Under the assumption that the time correlations can be neglected, i.e., 
\begin{equation*}
\langle \prod\limits_{k=0}^{n}N_j (t+k) \rangle = \prod\limits_{k=0}^n\langle N_j(t+k)\rangle = \langle N_j \rangle^{n+1},
\end{equation*}
the above expression simplifies considerably and can be expressed explicity in terms of the steady state density $\rho_{i+1}$ as:
\begin{align}\label{MFexpression}
r_i &= \begin{aligned}[t] &1-\rho_{i+1}(1-\omega_D)+ 2 \rho_{i+1}(1-\omega_D)[1-\rho_{i+1}(1-\omega_D)] \\
&+ 3 [\rho_{i+1}(1-\omega_D)]^2 [1-\rho_{i+1}(1-\omega_D)] + ... \nonumber\\
\end{aligned}\\
&= \sum_{n=1}^{\infty} n (\rho_{i+1}(1-\omega_D))^{n-1} (1-\rho_{i+1}(1-\omega_D)) \nonumber \\
&= \frac{1}{1-\rho_{i+1}(1-\omega_D)}.
\end{align}
The on-site residence time in eq.~\eqref{MFexpression} is the same as in eq.~\eqref{summation} under the identification of $ p_{i+1} \leftrightarrow 1-\rho_{i+1}(1-\omega_D)$. This reduces to $ p_{i+1} \leftrightarrow 1-\rho_{i+1}$ for the pure TASEP model, which is expected in the mean-field description \cite{cirillo2014residence}. Similarly the boundary condition on the rightmost site $N$ is given by 
\begin{equation}
r_{N} = \frac{1}{(1-\omega_D)\beta+\omega_D}.
\end{equation}

For the pure TASEP model, the total residence time $R_{\mathrm{\small{TASEP}}}$ can simply be written as
\begin{equation}\label{Rtasep}
R_{\mathrm{\small{TASEP}}} = \sum_{i=1}^{N} r_i.
\end{equation}
For the TASEP+LK model however, the possibility of a particle detaching before reaching the other end also needs to be taken into account. The modified equation for $R$ takes the following form:
\begin{equation}
R_{\mathrm{\small{TASEP}}} = r_1 + \sum_{i=2}^{N} r_i \prod_{k=2}^{i} f_{k-1,k} .
\label{probstay}
\end{equation}
For this modification, we consider the ensemble of steady state configurations that have a particle occupying the site $(i-1)$. The factor $f_{i-1,i}$ in eq.~\eqref{probstay}, defined for $2 \leq i \leq N $,  denotes the probability that a particle in this ensemble hops from site $(i-1)$ to the next site $i$ (does not undergo detachment). This probability is given by
\begin{widetext}
\begin{align}
f_{i-1,i} &= \big\langle \begin{aligned}[t] (1-\omega_D)[1-N_{i}(t)] &+ (1-\omega_D)^2 N_{i}(t)[1-N_{i}(t+1)]\\
&+(1-\omega_D)^3 N_{i}(t)N_{i}(t+1)[1-N_{i}(t+2)]+...\big\rangle \nonumber\\
\end{aligned}\\
&= \left\langle (1-\omega_D)(1-N_{i}(t)) + \sum_{j=1}^{\infty} (1-\omega_D)^{j-1} (1-N_{i} (t+j)) \prod_{k=0}^{j-1} N_{i} (t+k)\right\rangle.
\end{align} 
\end{widetext}
Applying the same mean-field approximation as in the derivation of eq. \eqref{MFexpression}, the equation for $f_{i-1,i}$ can be written as
\begin{align}\label{f_i}
f_{i-1,i} &= \sum_{n=1}^{\infty} \rho_{i}^{n-1}(1-\omega_D)^n (1-\rho_{i})\\
&= 1-\frac{\omega_D}{1-\rho_{i} (1-\omega_D)}.
\end{align}
To calculate the total residence time for the TASEP+LK model, we multiply the residence time of each site by the probability that a particle starting at the $\alpha$ boundary reaches this site (and thus does not detach before that). We thus write
\begin{widetext}
\begin{align}\label{RtasepLK}
R_{\mathrm{\small{TASEP+LK}}} &= r_1 + \left(1-\frac{\omega_D}{1-\rho_2 (1-\omega_D)}\right) r_2 +  \left(1-\frac{\omega_D}{1-\rho_2 (1-\omega_D)}\right)\left(1-\frac{\omega_D}{1-\rho_3 (1-\omega_D)}\right) r_3 + ... \nonumber \\
&= r_1 + \sum_{n=2}^{N}  r_n \prod_{k=2}^{n} \left(1-\frac{\omega_D}{1-\rho_k (1-\omega_D)}\right) = r_1 + \sum_{n=2}^{N}  r_n \prod_{k=2}^{n} (1-\omega_D r_k).
\end{align}
\end{widetext}
We see that eq.~\eqref{RtasepLK} reduces to eq.~\eqref{Rtasep} if we set $\omega_D$ = 0.

\subsection{Breakdown of mean-field in LD-HD phase}
\begin{figure}[h]
\includegraphics[width= 3.5 in]{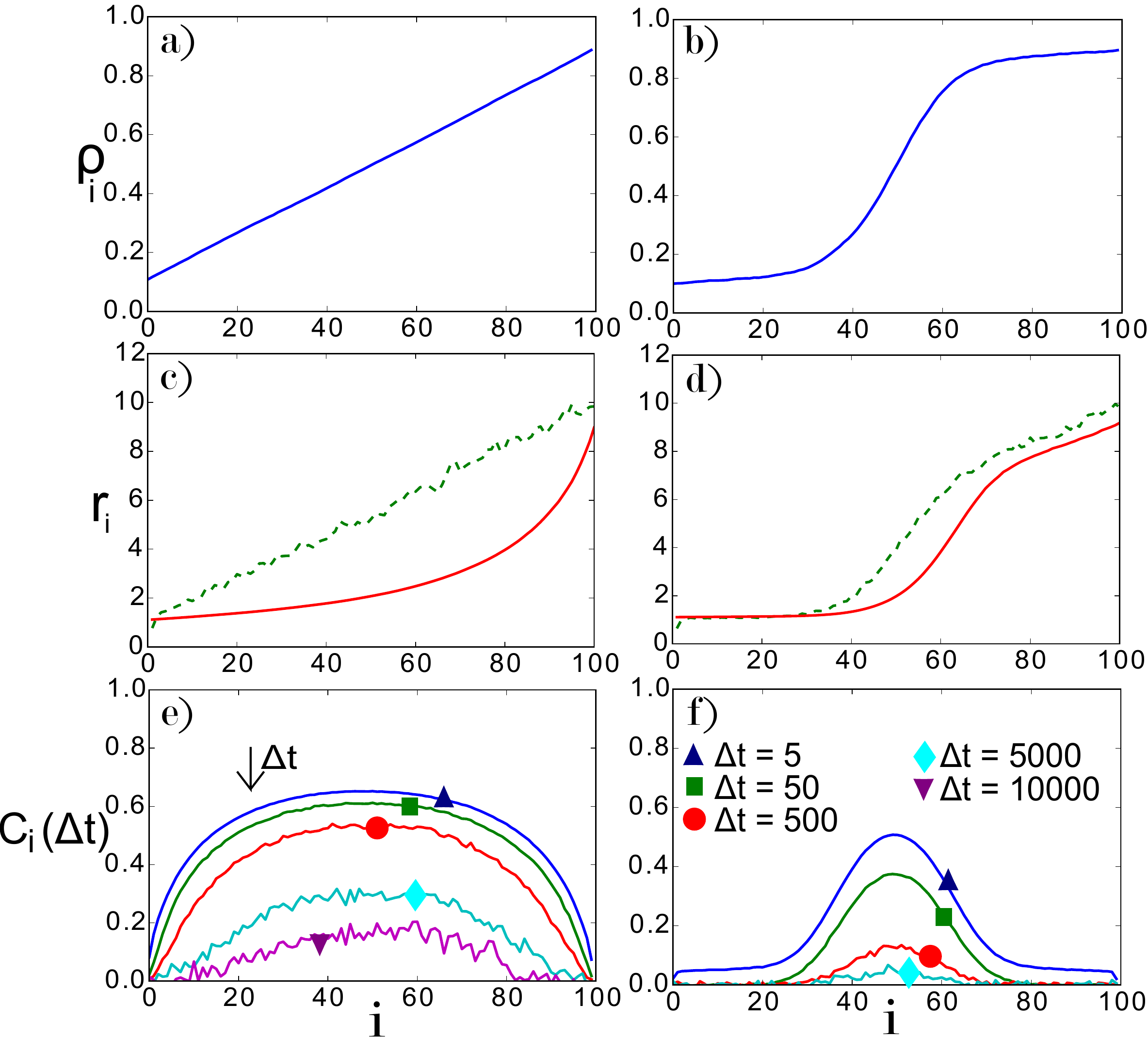}
\caption{Left panels correspond to pure TASEP and the right panels to TASEP + LK. (a) and (b): Steady state density profile. (c) and (d): On-site residence times from simulations (dashed lines) together with the mean-field prediction (solid lines). (e) and (f): Time correlation of local density at different time steps (see legend). Model parameters for TASEP + LK: $\alpha = \beta = 0.1$, $\Omega = 0.1$ and $N = 100$. TASEP model parameters: $\alpha = \beta = 0.1$ and $N = 100$.}
\label{timecorrs}
\end{figure}

Eqns \eqref{MFexpression}, \eqref{Rtasep} and \eqref{RtasepLK} are valid if the time correlations in local density are negligible. If, however, there are time correlations present in the local density, the mean-field expression underestimates the on-site residence times. Monte Carlo simulations revealed significant time correlations for systems in LD-HD phase, see fig.~\ref{timecorrs} e \& f. The time correlations $C_i (\Delta t)$ are calculated by

\begin{equation}
C_i (\Delta t) = \frac{\langle(N_i (t) -\rho_i) (N_i (t+\Delta t)-\rho_i)\rangle}{\langle (N_i (t) -\rho_i)^2 \rangle},
\end{equation}
where each Monte Carlo timestep in the random sequential update (for a description of this update procedure, see e.g. \cite{rajewsky1998asymmetric}) corresponds to a time interval of $\Delta t=1$.
The time correlations are rather long-lived in comparison to the mean-field estimate of residence time, especially in the neighborhood of the average shock location. We therefore expect eq.~\eqref{MFexpression} to give incorrect values for the on-site residence times for these profiles.  
In order to obtain accurate on-site residence time, a non-mean-field expression for $r_i$ for these profiles is thus needed.

\section{Non mean-field approach to residence time} \label{ourtheory}
A shock-type profile is characterized by the existence of a low-density phase at the $\alpha$ boundary and a high-density phase at the $\beta$ boundary. The sharp transition from one density phase to the other is considered as the shock interface. Two plots of the density profiles of such LD-HD phases are shown in Fig.~\ref{timecorrs}. In the TASEP model, the shock is completely delocalized. In the TASEP+LK model, the interface is localized only in the limit of $N\rightarrow \infty$. For finite $N$, the interface performs a random walk in a confining potential \cite{evans2003shock}. As shown above, time correlations in the local density are particularly significant near the average shock location persisting on time scales longer than the mean-field estimate. In order to derive an expression for the on-site residence times in presence of temporal correlations, we make the following assumptions about our system:
\begin{enumerate}
\item Each system site can at each timestep be determined to be either in the low density phase or in the high density phase.
\item A particle waiting to hop on site $i$ stays in the same density phase until it has hopped.
\end{enumerate}
In the following section, Sec.~\ref{flaglkrs}, we investigate the validity of these assumptions.

We start by rewriting eq.~\eqref{explicit} as 
\begin{equation}\label{taunew}
\tau_i (t) = \left[\Phi_i(t)+ (1-\Phi_i (t)) \right] \tau_i (t),
\end{equation}
where we define $\Phi_i(t)$ as a time-dependent \textit{phase constant}.  $\Phi_i(t)$ takes on value of $1$ if site $i$ belongs to the low-density phase and $0$ if site $i$ belongs to the high-density phase. 

We now take the average of both sides of this equation in the following way:
\begin{equation}
\langle \tau_i (t) \rangle = \langle \Phi_i(t)\tau_i (t) \rangle + \langle(1-\Phi_i (t)) \tau_i (t)\rangle.
\label{averageTau}
\end{equation}
Each of the two terms on the right hand side corresponds to an ensemble average. Using assumption 1, the fraction of systems in the ensemble that are in the low-density phase on site $i$ is given by $\langle \Phi_i(t) \rangle$. Similarly, the fraction of systems in the ensemble that are in the high-density phase on site $i$ is given by $\langle 1-\Phi_i(t) \rangle$. However, since $\Phi_i(t)$ is 1 in the low-density phase and 0 in the high-density phase, the term  $\langle \Phi_i(t)\tau_i (t) \rangle$ in eq.~\eqref{averageTau} gets a contribution only from those systems in the ensemble which are in low-density phase. Similary, $\langle(1-\Phi_i (t)) \tau_i (t)\rangle$ gets finite contribution only from the systems in high-density phase in the ensemble. We can thus write
\begin{align}
\langle \tau_i (t) \rangle = \langle \tau_i (t) \rangle^{\mathrm{\scriptscriptstyle{LD}}}_{i,t} \langle \Phi_i(t)\rangle + \langle \tau_i (t) \rangle^{\mathrm{\scriptscriptstyle{HD}}}_{i,t} \langle 1-\Phi_i(t)\rangle,
\end{align}

where $\langle ... \rangle^{\mathrm{\scriptscriptstyle{LD}}}_{i,t}$ and $\langle ... \rangle^{\mathrm{\scriptscriptstyle{HD}}}_{i,t}$ denote an average over the systems in LD phase and the systems in HD phase at site $i$ at time $t$ respectively.

We now have a weighted sum of the average residence time in the low-density phase and the average residence time in the high-density phase. In a recent study~\cite{cirillo2014residence} it has been shown that the mean-field prediction for the residence times is accurate for a purely HD system as well as a purely LD system. We now make use of assumption 2 above, which allows us to identify $\langle \tau_i (t) \rangle^{\mathrm{\scriptscriptstyle{LD}}}_{i,t}$ with the on-site residence time on site $i$ of a pure LD system, and $\langle \tau_i (t) \rangle^{\mathrm{\scriptscriptstyle{HD}}}_{i,t}$ with the on-site residence time on site $i$ of a pure HD system. We thus arrive at the following expression for the on-site residence time for an LD-HD system: 
\begin{equation} \label{rnew}
r_i = \frac{\lambda_i}{1-\rho_{\mathrm{LD}, i} (1-\omega_D)} + \frac{1-\lambda_i}{1-\rho_{\mathrm{HD}, i}(1-\omega_D)}.
\end{equation}
Here, we have identified $\langle \Phi_i \rangle = \lambda_i$.

As our parameter $\lambda_i$ is equal to the fraction of time that site $i$ is part of the low-density phase, and $1-\lambda_i$ is equal to the fraction of time site $i$ is in the high density phase, we can write the following expression for the average density profile:
\begin{equation}\label{rhonew}
\rho_i = \lambda_i \rho_{\mathrm{LD},i} + (1-\lambda_i ) \rho_{\mathrm{HD},i}.
\end{equation}
From this, we can obtain an expression for $\lambda_i$:
\begin{equation}\label{lambda}
\lambda_i = \frac{\rho_i - \rho_{\mathrm{HD},i}}{\rho_{\mathrm{LD},i} - \rho_{\mathrm{HD},i}}.
\end{equation}
This equation, together with eq.~\eqref{rnew}, gives us a completely analytical result for the on-site residence times in terms of $\rho_{\mathrm{LD},i}$, $\rho_{\mathrm{HD},i}$ and $\rho_i$.

Eq. \eqref{Rtasep} for the total residence time $R$ still holds for pure TASEP in LD-HD phase. For TASEP+LK however, eq.~\eqref{RtasepLK} is no longer valid as eq.~\eqref{f_i} ignores time correlations. A non mean-field expression for $f_{i-1,i}$ for systems in LD-HD phase can be derived using the same approach as in the derivation of eq.~\eqref{rnew}. The result is:
\begin{align}
f_{i-1,i} = &1 - \frac{\omega_D \lambda_i}{1-\rho_{\mathrm{LD},i} (1-\omega_D)} \nonumber \\
&- \frac{\omega_D (1-\lambda_i)}{1-\rho_{\mathrm{HD},i} (1-\omega_D)}.
\end{align}
The total residence time $R$ on the lattice is then given by
\begin{align}\label{RtasepLKnonMF}
R_{TASEP+LK} &= \begin{aligned}[t] r_1 &+ \sum_{n=2}^{N}  r_n \prod_{k=2}^{n} \left( 1 -\frac{\omega_D \lambda_k}{1-\rho_{\mathrm{LD},k} (1-\omega_D)} \right. \nonumber \\
&- \left. \frac{\omega_D (1-\lambda_k)}{1-\rho_{\mathrm{HD},k} (1-\omega_D)} \right) \nonumber \\
\end{aligned} \\
&= r_1 + \sum_{n=2}^{N}  r_n \prod_{k=2}^{n} (1-\omega_D r_k).
\end{align}
This equation is identical to eq.~\eqref{RtasepLK}, but with $r_i$ now given by eq.~\eqref{rnew}.

\section{Domain wall dynamics} \label{DWT}
\subsection{Flag theory for TASEP+LK with random sequential update}\label{flaglkrs}
In the previous section, we saw there are two conditions that need to be satisfied in order for eq.~\eqref{rnew} to hold: (1) a location can be assigned to the shock at each timestep and (2) the shock moves in such a way that a particle crosses the shock only when it makes a hop. In order to investigate if these conditions are indeed satisfied, a description of the shock between the low density phase and the high density phase is thus needed. 

For the pure TASEP model, a microscopic description of the shock location has been derived in Ref.~\cite{ferrari1994shock}. This was done by introducing a \textit{second class particle}. This is a particle that hops like the system particles if it has a hole to its right, but if a system particle to its left tries to hop, the particles switch places. In Ref.~\cite{ferrari1994shock}, it was proven that this second class particle indeed tracks the location of the shock. 
This implies that for the pure TASEP model, our two conditions are indeed satisfied: all sites can be determined to be either in the high or the low density phase, and a particle makes the transition between phases only when it hops onto the site where the second-class particle is located.

We would like to extend the idea of this shock marker to a TASEP+LK system. A problem for doing this is that this second class particle would have to be subject to langmuir kinetics, allowing it to detach, and thus removing the shock marker from the system.

A solution to this is to consider a \textit{flag} as described in Ref.~\cite{cividini2014exact} instead of the second class particle. This flag was introduced in Ref.~\cite{cividini2014exact} to track the location of the shock in pure TASEP with parallel update, and obeys the following motion rules:
\begin{enumerate}
\item The flag is first placed at the leftmost particle of the leftmost cluster
\item If the particle with the flag on it moves, the flag moves with it
\item If a particle is blocked by the particle carrying the flag, the flag is transferred to this blocked particle.
\end{enumerate}  
If we apply these motion rules to a TASEP system with random sequential update, we see that the motion of the flag is identical to the motion of the second class particle.

\begin{figure}
\includegraphics[width = \columnwidth]{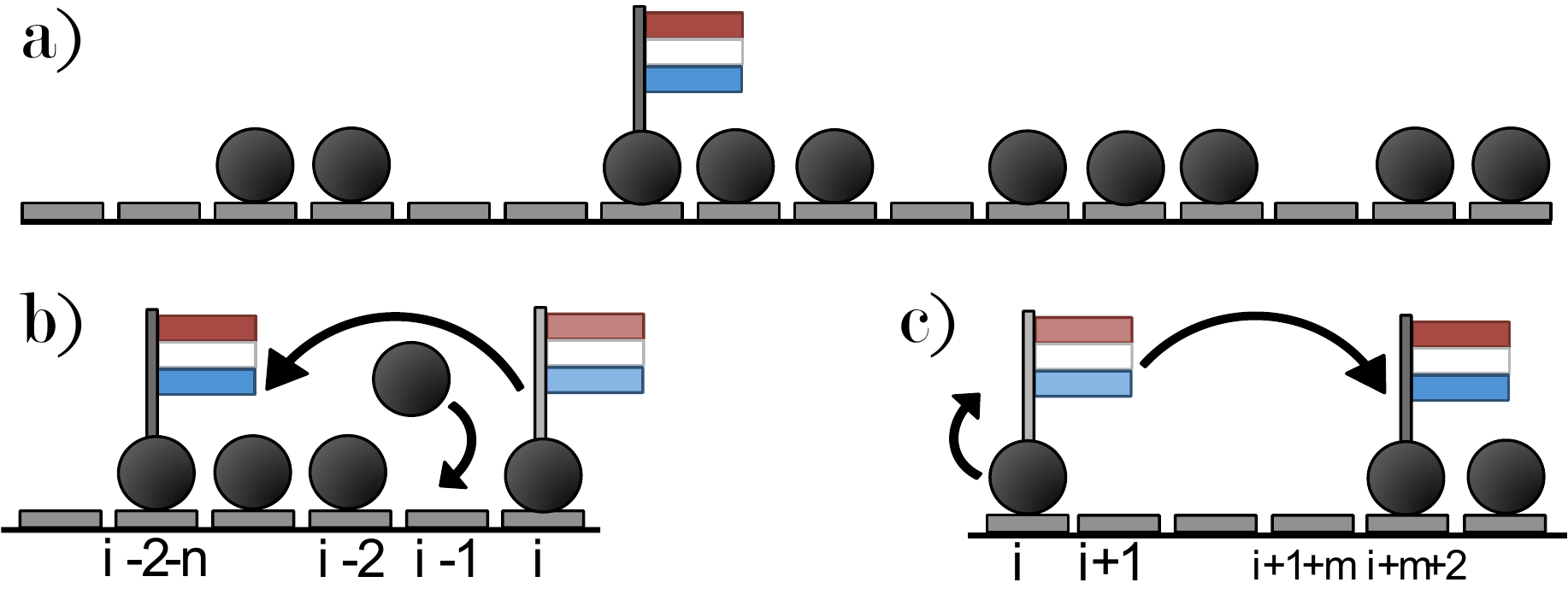}
\caption{A graphic representation of flag theory. (a) A flag is placed on a system particle to keep track of the shock location. Its movement rules are such that the flag is always located on the leftmost particle of a cluster. (b) and (c) TASEP + LK:  If two clusters merge forming a single cluster due to an attachment event, the flag which was previously on the right cluster, moves to the leftmost particle of the new merged cluster. In the event of detachment of the particle carrying the flag, the flag is transferred to the leftmost particle of the cluster immediately to the right of the detaching particle.}
\label{flags}
\end{figure}

A note on the application of rule 1 in our model: due to clusters spontaneously forming in the low density region, the flag is not necessarily placed close to the shock interface using this rule. However, it ensures that the flag is placed on particle such that the flag has no particle to the left of it. After a few timesteps, the flag will move close to the shock interface due to rules 2 and 3 above and will be a good indicator of the shock position.  

For the TASEP+LK and TASEP+MILK cases, we have additional movement rules for the flag due to attachment and detachment events. We propose the following additions (shown graphically in Fig.~\ref{flags}): 
\begin{enumerate}
\setcounter{enumi}{3}
\item If the flag is located at site $i$, a particle attaches at site $i-1$ and sites $i-2-n$ through $i-2$ are occupied, the flag moves to site $i-2-n$. If site $i-2$ is unoccupied, the flag moves to site $i-1$.
\item If the flag is located at site $i$, the particle at site $i$ detaches and sites $i+1$ through $i+1+m$ are unoccupied, the flag moves to site $i+1+m+1$. If site $i+1$ is occupied, the flag moves to site $i+1$.
\end{enumerate}
These rules are natural if an attachment event between the flag and a cluster ending two places to the left of it is viewed as several particles coming in at the same timestep. Similarly, a detachment event of the particle which carries the flag and has several holes to its right is viewed as several holes coming in at the same timestep.

\begin{figure}
\includegraphics[width = \columnwidth]{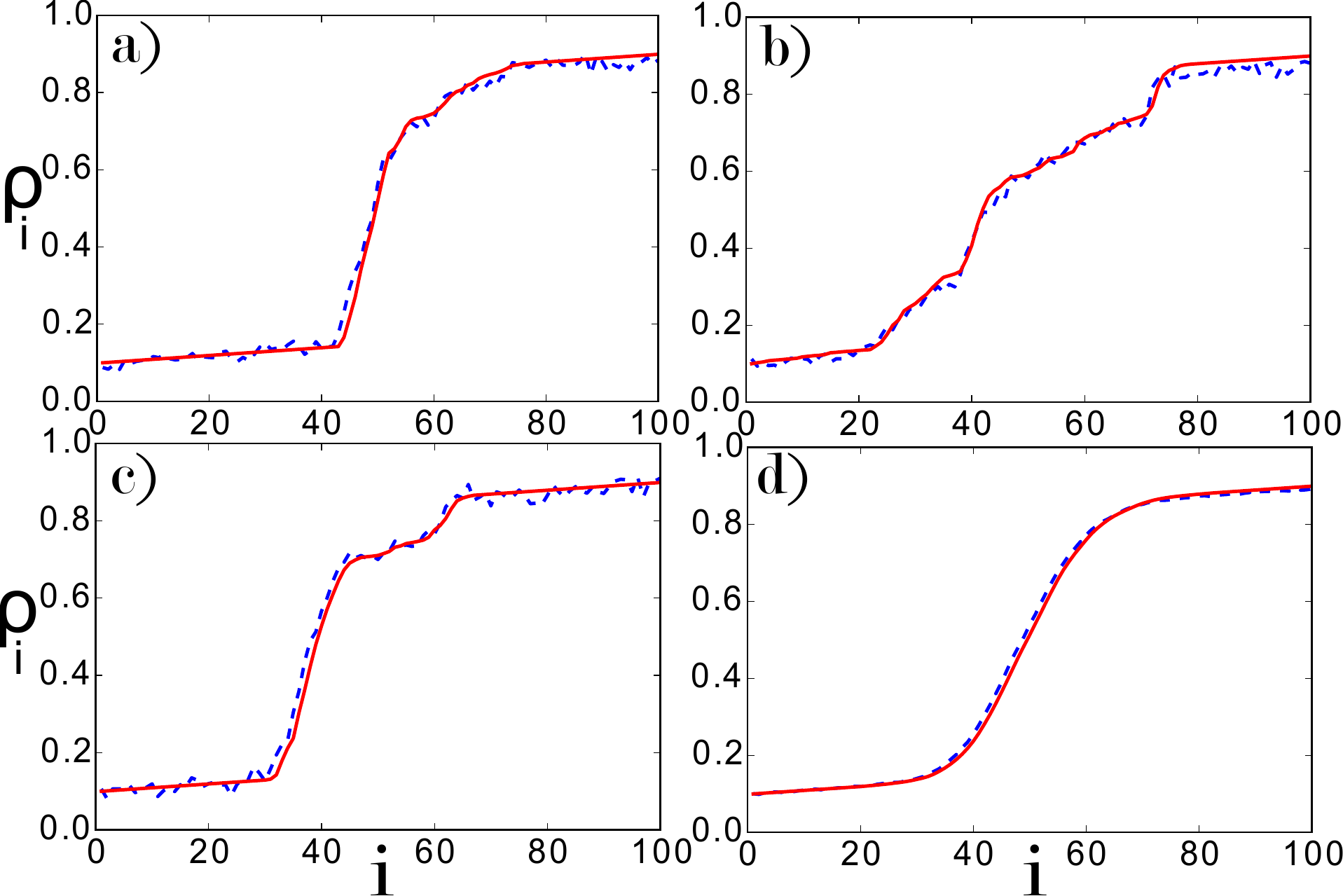}
\caption{Average density profiles for TASEP+LK obtained by keeping track of the shock location using our rules for the flag movement. The solid red line corresponds to the profile predicted from the flag locations, the blue dashed line corresponds to the actual density profile obtained in the same simulation. The density profiles of figures a), b) and c) are taken over a relatively short timescale (600 timesteps in the Monte Carlo simulation), the profile of figure d) is taken over a long timescale (50000 timesteps).
The excellent agreement between the predicted and numerically obtained profiles in (d) is a clear indication of the accurate recovery of shock position distribution in the steady state. 
Model parameters: $\alpha =\beta =\Omega = 0.1$, $N=100$.}
\label{flagprofiles}
\end{figure}

To test the validity of this description of the flag location, we performed simulations keeping track of the flag locations together with the average density profile. The distribution of flag locations was used together with the LD and HD average density solutions eqs.~\eqref{soln1} and \eqref{soln2} to predict the average density at each site. Density profiles obtained this way matched very well with the actual average density, for short as well as for longer run times, as shown in Fig.~\ref{flagprofiles}. With the above modifications, flag theory can thus be applied to TASEP+LK as well.

With these additions, it is now possible for a particle to switch phases while waiting to hop. The derivation in the previous section is thus no longer strictly valid. It is however still a good approximation if such a switch is a rare event. We therefore consider the fraction $\chi$ of flag movements due to attachment and detachment events. For a flag located at site $i$, this fraction is given by
\begin{align}
\chi &= \langle N_{i-2} (1-N_{i-1}) \omega_A + N_i \omega_D (1-N_{i+1})\rangle \nonumber \\
&= \omega_A \langle N_{i-2} (1-N_{i-1}) \rangle + \omega_D \langle N_{i-2} (1-N_{i-1}) \rangle.
\end{align}

We see that this fraction is much smaller than 1 in the case that $\omega_A, \omega_D \ll 1$. This is indeed the case for LD-HD systems as long as the number of sites $N$ is not of order 1. To see this, we note that for the TASEP+LK model, we have a shock in the case that $\Omega < 1-\beta-\alpha$. This implies that $\omega_A, \omega_D < \frac{1-\beta - \alpha}{N}$, so $\omega_A, \omega_D$ can only be of order 1 if $N$ is of order 1. 

We conclude that the two assumptions for the derivation of the previous section are indeed valid. It is possible to assign a location to the shock at each timestep, thus it is possible to determine each system site to be either in the low density phase or in the high density phase at each timestep. As long as the number of sites $N$ is not of order 1, a particle, to a good approximation only switches phases while performing a hop. Thus it is indeed justified to write eq.~\eqref{rnew} for the on-site residence times for LD-HD systems.

\subsection{Density profiles in LD-HD phase}\label{densityshock} 
In section \ref{ourtheory}, eq.~\ref{rnew} for the on-site residence times for an LD-HD system was derived in terms of $\rho_{\mathrm{LD}}(x)$, $\rho_{\mathrm{HD}}(x)$ and $\rho (x)$. In order to calculate on-site residence times from this equation, an equation for $\rho (x)$ is thus needed. A derivation for this has been done in Ref.~\cite{evans2003shock} using Domain wall theory (DWT). We briefly summarize their results below.

The average density profile in this system (written in the continuum limit) is described by
\begin{equation}\label{rhofromp}
\rho (x) = \rho_{\mathrm{HD}} (x) \int_0^x p(x') dx' + \rho_{\mathrm{LD}} (x) \int_x^1 p(x')dx'.
\end{equation}
In this equation, $p(x)$ describes the distribution of the locations of the shock. The system length $L$ has been set to 1. The solution for $p(x)$ found by \cite{evans2003shock} is given by 
\begin{equation} \label{p}
p(x) = \frac{1}{\mathcal{N} \omega_l (x)} e^{-N \int_{x_0}^{x} (1-\frac{\omega_r(x')}{\omega_l(x')})dx'}.
\end{equation}
Here, $N$ is the number of sites on the lattice, and $\omega_r$ and $\omega_l$ are the hopping rates of the shock location to the right and left respectively as shown in Ref.~\cite{evans2003shock}
\begin{align*}
&\omega_r = \frac{\rho_{\mathrm{HD}}}{\rho_{\mathrm{HD}} - \rho_{\mathrm{LD}}} \\
&\omega_l = \frac{\rho_{\mathrm{LD}}}{\rho_{\mathrm{HD}} - \rho_{\mathrm{LD}}}.
\end{align*}
The normalization constant $\mathcal{N}$ is chosen such that $\int_0^1 p(x) dx = 1$. The term $x_0$ can be chosen arbitrarily as it will be absorbed in the normalization constant $\mathcal{N}$.

In the following sections, we make use of these results together with our eqs. \eqref{rnew} and \eqref{lambda} to obtain residence time predictions in specific conditions.

\section{Results}\label{results}
\subsection{Shock-type profiles in the TASEP model} \label{TASEP}
For the pure TASEP case, we have a completely delocalized shock in the case that $\alpha = \beta < \frac{1}{2}$. Our two solutions  for the differential equation \eqref{diffLK} that meet a boundary condition are given by eqs. \eqref{soln1} and \eqref{soln2} with $\Omega = 0$.

We use eq.~\eqref{p} to calculate $p(x)$. Since $\omega_r = \omega_l$ for the pure TASEP model, this expression reduces to
\begin{equation}
p_{\mathrm{\scriptscriptstyle{TASEP}}}(x) =  1.
\end{equation}
We see here that our distribution function is no longer a function of the number of sites $N$. This means that the shape of our profile is now size-independent. \\
Using eq.~\eqref{rhofromp} in the continuum limit, the expression for the average density profile $\rho (x)$ becomes
\begin{equation}
\rho (x) = \alpha + x (1 -2 \alpha),
\end{equation}
which agrees with the result in Ref.~\cite{derrida1993exact}.

\begin{figure}
\includegraphics[width = \columnwidth]{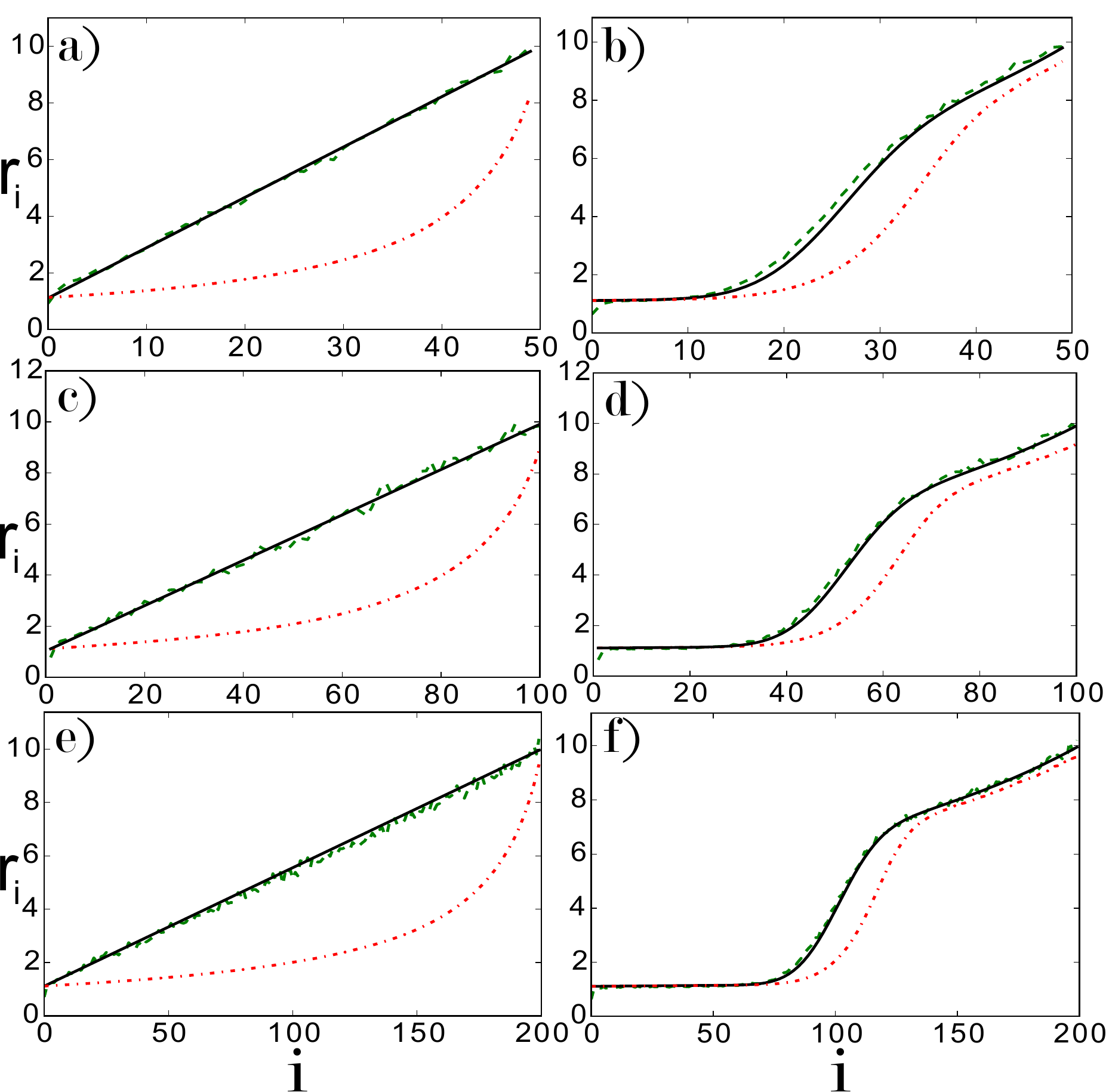}
\caption{On-site residence times for different system lengths. The left panels correspond to pure TASEP with $\alpha = \beta = 0.1$, the right panels correspond to TASEP+LK with $\alpha = \beta = \Omega = 0.1$.  Figures (a) and (b) correspond to $N=50$, (c) and (d) to $N=100$, (e) and (f) to $N=200$. Green dashed lines: on-site residence times obtained from simulations. Black solid lines: analytical on-site residence times from eq.~\eqref{rnew}. Red dash-dotted lines: mean-field result for residence times.}
\label{taseprt}
\end{figure}

Using eq.~\eqref{rnew} to obtain the on-site residence time we obtain
\begin{equation}\label{rtasep_onsite}
r(x) = \frac{\alpha + x (1-2\alpha)}{\alpha (1-\alpha)}.
\end{equation}
The results from Monte Carlo simulation together with the prediction by eq.~\eqref{rtasep_onsite} are shown in Fig.~\ref{taseprt}. The total residence time on the lattice is now given by 
\begin{equation}\label{RtotalTasep}
R = N \int_0^1 r(x) dx = N \frac{1}{2 \alpha (1-\alpha)}.
\end{equation}
We thus recover the linear dependence of the residence time on the system length, in accordance with the results of Ref.~\cite{cirillo2014residence}. The results for our numerical simulation compared to the results predicted by eq.~\eqref{RtotalTasep} are presented in Fig.~\ref{allrs}.

\subsection{Shock-type profiles in the TASEP+LK model}
The derivation of the analytical profile for a shock-type profile in the TASEP+LK model has been done explicity in Ref.~\cite{evans2003shock} for the case that $\Omega_A = \Omega_D = \Omega$. The result is:
\begin{align} \label{rholk}
\rho (x) = &\frac{\Delta}{2} \left[1+\erf\left(2 \sqrt{\frac{N \Omega \Delta}{(1+\Delta) (1-\Delta)}} (x-x_s)\right) \right] \nonumber \\
&+ \Omega x + \alpha,
\end{align}
where $\Delta$ is the height of the shock, given by $\Delta = \rho_{\mathrm{HD}} - \rho_{\mathrm{LD}}$, and $x_s$ is the average shock position.
The on-site residence time can now be calculated using eq.~\eqref{rnew} in the continuum limit:

\begin{align}\label{rofx}
r(x) = &\frac{\lambda (x)}{1-(\Omega x + \alpha)(1-\omega_D)} \nonumber \\
&+ \frac{1-\lambda (x)}{1-(\Omega (x-1) + 1 - \beta)(1-\omega_D)}, 
\end{align}
where $\lambda (x)$ is given by
\begin{align}
\lambda (x) = \frac{\frac{\Delta}{2} \left[1+\erf\left(2 \sqrt{\frac{N \Omega \Delta}{(1+\Delta) (1-\Delta)}} (x-x_s)\right) \right]}{1+\alpha-\Omega-\beta} + 1.
\end{align}

The results for our numerical simulation compared to the results predicted by eq.~\eqref{rofx} are presented in Fig.~\ref{taseprt}. 

\begin{figure}
\includegraphics[width = \columnwidth]{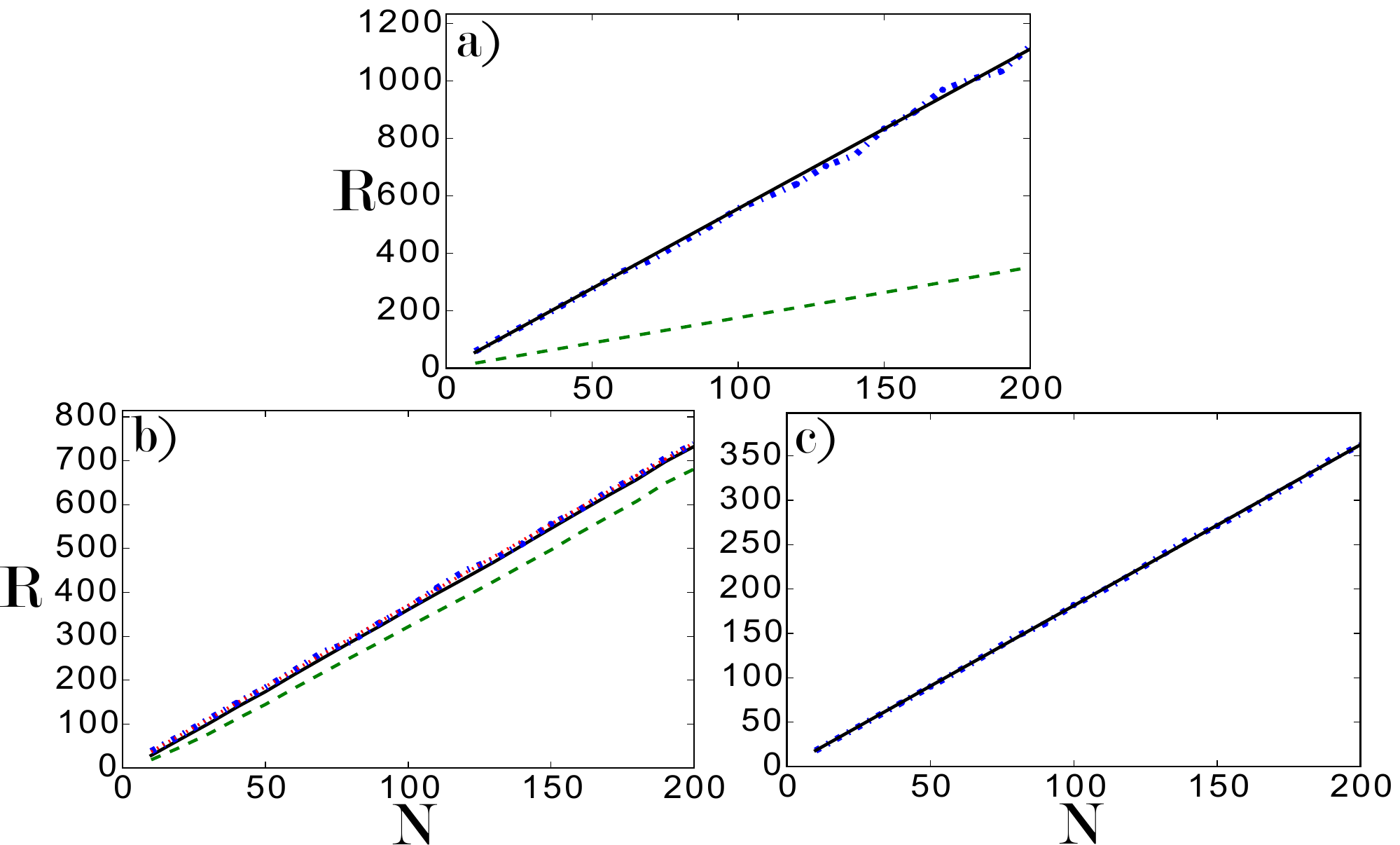}
\caption{Total residence times $R$ for TASEP in LD-HD phase (a), for TASEP+LK in LD-HD phase (b) and for TASEP+LK in MC phase (c). Blue dash-dotted line: results from Monte Carlo simulation. Black line: non mean-field prediction. Green longitudinally-dashed line: mean-field prediction. Red transversally-dashed line: prediction by eq.~\eqref{Rlinear}. Panel (a): non mean-field result calculated with eq.~\eqref{Rtasep}, model parameters $\alpha = \beta = 0.1$. Panel (b): non mean-field result calculated with eq.~\eqref{RtasepLKnonMF}, model parameters $\alpha = \beta = \Omega = 0.1$. Panel (c): analytical result calculated with eq.~\eqref{RMC}, model parameters $\alpha=\beta=0.5$, $\Omega = 0.1$.}
\label{allrs}
\end{figure}

The total residence time $R$ can be obtained numerically using eq.~\eqref{RtasepLKnonMF} with eq.~\eqref{rofx} as arguments. The mean-field result would be given by evaluating eq.~\eqref{RtasepLK} with eq.~\eqref{MFexpression} as arguments. The results of monte-carlo simulations are shown in Fig.~\ref{allrs}.
For large $N$, the $N$-dependence of $R$ is very close to linear as can be seen in Fig.~\ref{allrs}. An explicit expression for $R$ in the large $N$ limit is given in appendix \ref{linearR}, eq.~\eqref{Rlinear}. Since $R \sim N$, it follows that one can define a transport coefficient analogous to mobility (a material property independent of system size) in the limit of large $N$ for TASEP as well as TASEP + LK.

Until now, we have focused on systems with boundary conditions chosen such that the steady state density profile is described as the LD-HD phase. In the following section, we consider systems in MC phase and demonstrate that our theoretical approach is equally applicable to these systems as well.
 
\subsection{Residence time for TASEP+LK in maximum current phase}
The residence time of a particle when the system is in MC phase has been studied in Ref.~\cite{cirillo2014residence}. It was shown that the total residence time scales linearly with the system size. However, the study focused only on pure TASEP. We will now show that this linear scaling remains present for a TASEP+LK system as well.

In a maximum current phase, characterized by a constant density in the bulk, depending on the mismatch between the boundary conditions and the Langmuir isotherm, boundary layers can exist. The width of the boundary layers relative to the system size approaches zero in the limit of $N\rightarrow \infty$. Below we assume that the boundary layers can be neglected in calculating the total residence time.

An exact expression for the total residence time for a constant density can be obtained by inserting $\rho_i = \bar{\rho}$ in eq.~\eqref{RtasepLKnonMF} which yields
\begin{equation} \label{Rconstantdensity}
R = \frac{ \left[ (N-\Omega)(1-\bar{\rho})\left(1-\left[1- \frac{\Omega}{\Omega \rho -\rho N + N} \right]^N\right)\right]}{\Omega (1-\bar{\rho} (1-\frac{\Omega}{N}))}.
\end{equation}

It can be easily shown that for $N \gg 1$, eq.~\eqref{Rconstantdensity} reduces to 
\begin{equation}\label{RMC}
R = \left( 1- \mathrm{e}^{-\frac{\Omega}{1-\bar{\rho}}} \right)N/\Omega. 
\end{equation}
It follows that for a constant bulk density, the total residence time scales linearly with the system size. The factor $\mathrm{e}^{-\frac{\Omega}{1-\bar{\rho}}}$ is the fraction of particles which starting on the first site of the lattice hop all the  way to the $\beta$ boundary. This factor is 1 in the limit of $\Omega/(1-\bar{\rho}) \ll 1$ in which case, the total residence time scales as $R \approx N/(1 - \bar{\rho})$. On the other hand, when $\Omega/(1-\bar{\rho}) \gg 1$, almost every particle detaches from the lattice before reaching the $\beta$ boundary. In that case, the total residence time, consistent with the assumed mesoscopic scaling is $R \approx 1/\omega_D = N/\Omega$.

\section{Residence time of a probe particle}\label{MILKsection}
Including mutual interactions between particles in addition to the hard-core repulsion gives rise to several interesting features. In the Katz-Lebowitz-Spohn (KLS) model, the hopping rates are modified depending on the occupancy of the next-nearest neighbor resulting in additional correlations~\cite{katz34nonequilibrium}. This model gives rise to exotic features such as localized downwards shocks and phase separation into three distinct regimes~\cite{popkov2003localization}. In the model TASEP + MILK shown schematically in Fig.~\ref{MILKschematic}, the attachment and detachment rates are modified depending on the occupancy of the adjacent sites~\cite{vuijk2015driven}. These additional interactions increase (decrease) the effects of boundaries on the phase behavior of the model. For such models with additional interactions, it is in principle possible to calculate the on-site residence times. However, a more  general and considerably simple description of residence time is possible by introducing a probe particle. The probe particle interacts via hard-core repulsion and does not undergo detachment. We consider a system consisting of of a single probe particle and all the other particles following TASEP (with or without LK). It is interesting to study the residence time of a single such probe particle. 

\subsection{Probe particle dynamics}
The residence time of the probe particle can be calculated using the same approach as in the derivation of eq. \eqref{explicit} for $r_i$ in the TASEP+LK case. The difference is now that the probe particle cannot detach. As the $\omega_D$-factors take the detachment of the particle considered into account, the on-site residence time of the probe particle is obtained by setting $\omega_D = 0$ in eq.~\eqref{explicit}.

Note that the resulting expression is a general one, valid for a probe particle in the TASEP model with any type of further modified interactions. The modifications of the interactions will all be captured in the average occupations and the time correlation functions. If time correlations can be ignored, this expression reduces to

\begin{equation}
r_i = \frac{1}{1-\rho_i}.
\end{equation}

For any system in LD-HD phase subject to the two assumptions stated in section \ref{ourtheory}, the on-site residence time is given by

\begin{equation}\label{rshockprobe}
r_i = \frac{\lambda_i}{1-\rho_{\mathrm{LD}}} + \frac{1-\lambda_i}{1-\rho_{\mathrm{HD}}}.
\end{equation}
The derivation of this equation is completely analogous to the derivation in section \ref{ourtheory}.

We now apply the description of the residence times of a probe particle to a model with modified interactions recently proposed in \cite{vuijk2015driven}: the totally asymmetric exclusion process together with mutually interactive Langmuir kinetics.

\subsection{Results for a probe particle in a TASEP+MILK system}
\begin{figure}[h]
\includegraphics[width = \columnwidth]{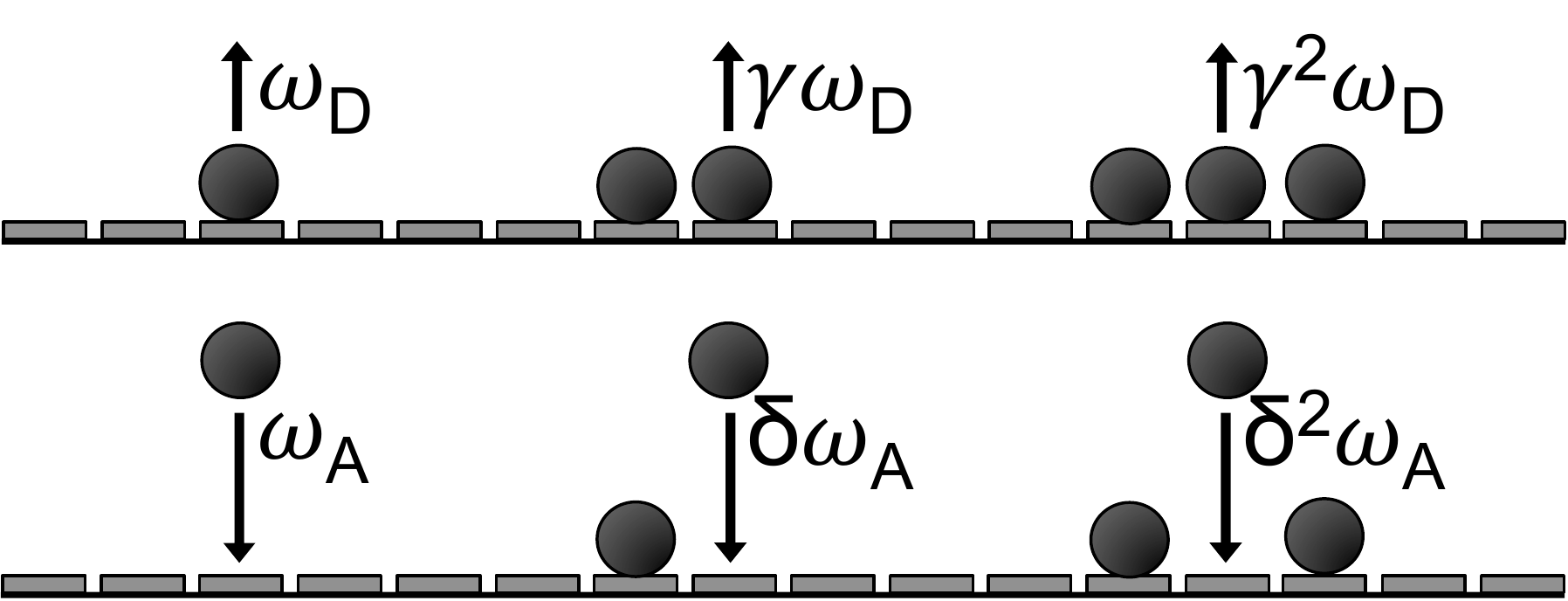}
\caption{A graphic representation of the TASEP+MILK model. Every particle performs ordinary TASEP in addition to Langmuir kinetics which depend on the ocupation states of neighboring sites. The detachment rate of a particle is scaled by a factor $\gamma$ if one of its neighboring sites is occupied. If both neighboring sites are occupied then the rate is scaled by a factor of $\gamma^2$. Similarly, the attachment rate is scaled by $\delta$ when only one of the neighboring sites is occupied and by $\delta^2$ when both are occupied.}
\label{MILKschematic}
\end{figure}

A schematic of TASEP + MILK system is shown in Fig.~\ref{MILKschematic}. Every particle is subjected to the same rules as in TASEP. In addition, every particle can undergo attachment and detachment as in Langmuir kinetics with the modification that the kinetic rates are dependent on the state of neighboring sites. If a particle has one neighbouring particle, its detachment rate becomes $\gamma \omega_D$. If it has two neighbouring particles, its detachment rate becomes $\gamma^2 \omega_D$. If there are no neighbouring particles, its detachment rate remains $\omega_D$. Similarly, the attachment rate becomes $\delta \omega_A$ or $\delta^2 \omega_A$ if one or both neighboring sites are occupied, respectively. If there are no neighbouring particles, its attachment rate will remain equal to $\omega_A$.

The differential equation describing the steady-state solutions density profile to first order in $\epsilon$ is given by \cite{vuijk2015driven}
\begin{multline}\label{diffMILK}
\epsilon \partial_x^2 \rho + (2\rho-1)\partial_x \rho  + \Omega_A [1+\rho(\delta-1)]^2 (1-\rho) \\- \Omega_D [1 + \rho (\gamma-1)]^2 \rho = 0.
\end{multline}
This equation has not been solved in general, but solutions are known for certain parameter regimes \cite{vuijk2015driven}.

In the TASEP+MILK model, shock-type profiles have been found to occur as well \cite{vuijk2015driven}. At the end of this section we apply our method to determine the residence time for these profiles, and compare with simulations. 

\begin{figure}
\includegraphics[width = \columnwidth]{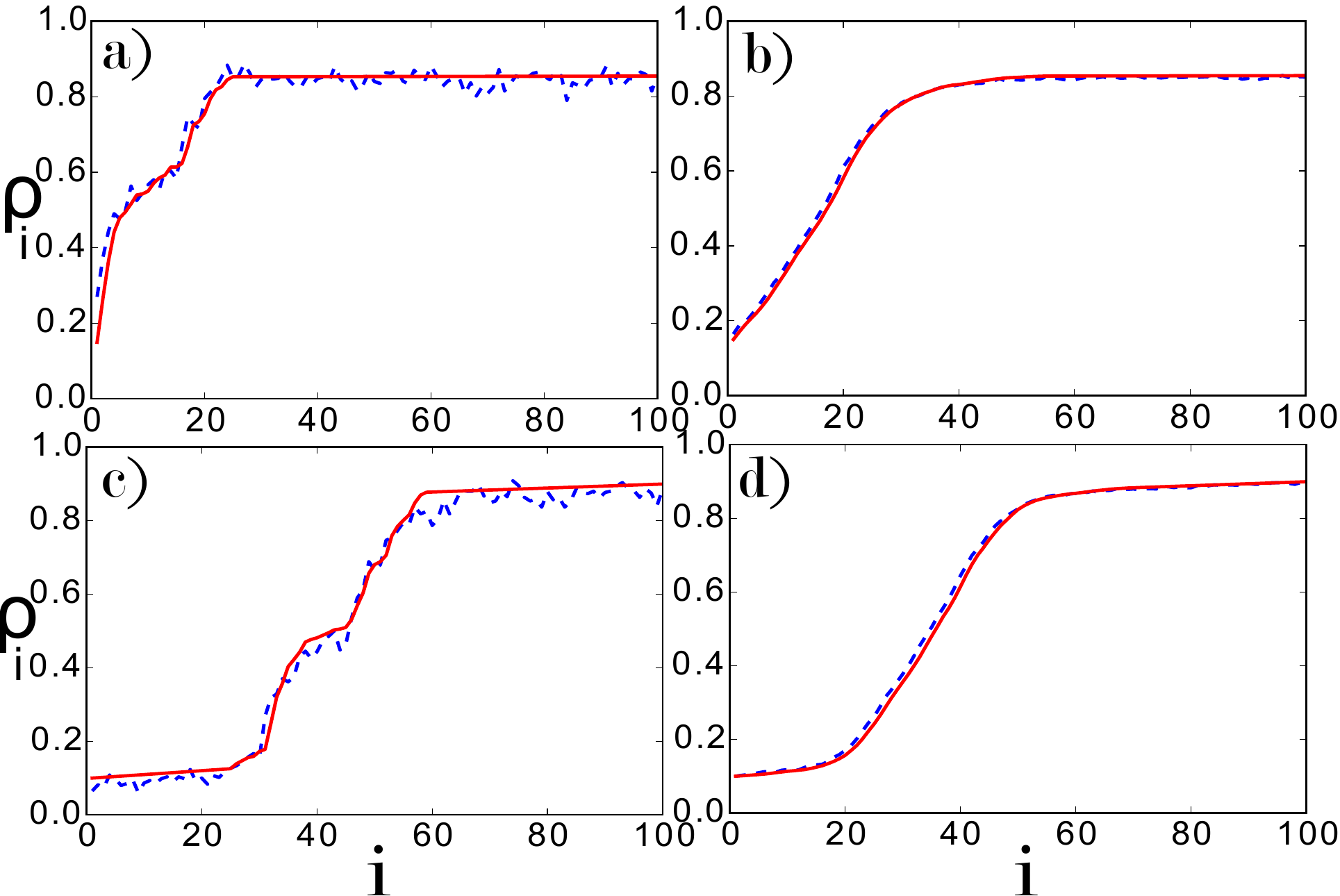}
\caption{Average density profiles for TASEP+MILK obtained by keeping track of the shock location using our rules for the flag movement. The solid red line corresponds to the profile predicted from the flag locations, the blue dashed line corresponds to the actual density profile obtained in the same simulation. The density profiles of figures a) and c) are taken over a relatively short timescale (600 timesteps in the Monte Carlo simulation), the profiles of figures b) and d) are taken over a long timescale (50000 timesteps). Model parameters for a) and b): $\delta = \gamma = 0$, $\alpha = \beta = 0.15$, $\Omega = 0.1$, $N=100$. Model parameters for c) and d): $\delta = 1 + \psi$ and $\gamma = 1-\psi$ with $\psi = 0.2$, $\alpha = 0.05$, $\beta = 0.15$, $\Omega = 0.3$, $N=100$. As in Fig.~\ref{flagprofiles}, the excellent agreement between the predicted and numerically obtained profiles in b) and d) is again a clear indication of the accurate recovery of shock position distribution in the steady state.}
\label{milkflagprofiles}
\end{figure}

First we investigate if it is justified to use the assumptions stated in section \ref{ourtheory} to the TASEP+MILK model. For this model, we can still use the rules for flag movement in a TASEP+LK stated in section \ref{flaglkrs}. This is because there are no new ways for particles to hop or attach to sites in the TASEP+MILK model; the only modification is in the probability of these events. We can calculate the fraction of flag movements due to attachment and detachment events using the same approach as in section \ref{flaglkrs}, the result is

\begin{equation}
\chi = \gamma^2 \omega_A \langle N_{i-2} (1-N_{i-1}) \rangle + \delta \omega_D \langle N_{i-2} (1-N_{i-1}) \rangle.
\end{equation}
We see that this fraction is much smaller than 1 as long as $\gamma^2 \omega_A  \ll 1$ and $\delta \omega_D \ll 1$. To test the validity of this description of the flag location in TASEP+MILK, we again performed simulations keeping track of the flag locations together with the average density profile. The approach is the same as for the plots in Fig.~\ref{milkflagprofiles}. Density profiles obtained this way matched very well with the actual average densities, for short as well as for longer run times, as shown in Fig.~\ref{milkflagprofiles}.

\begin{figure}[h]
\includegraphics[width = \columnwidth]{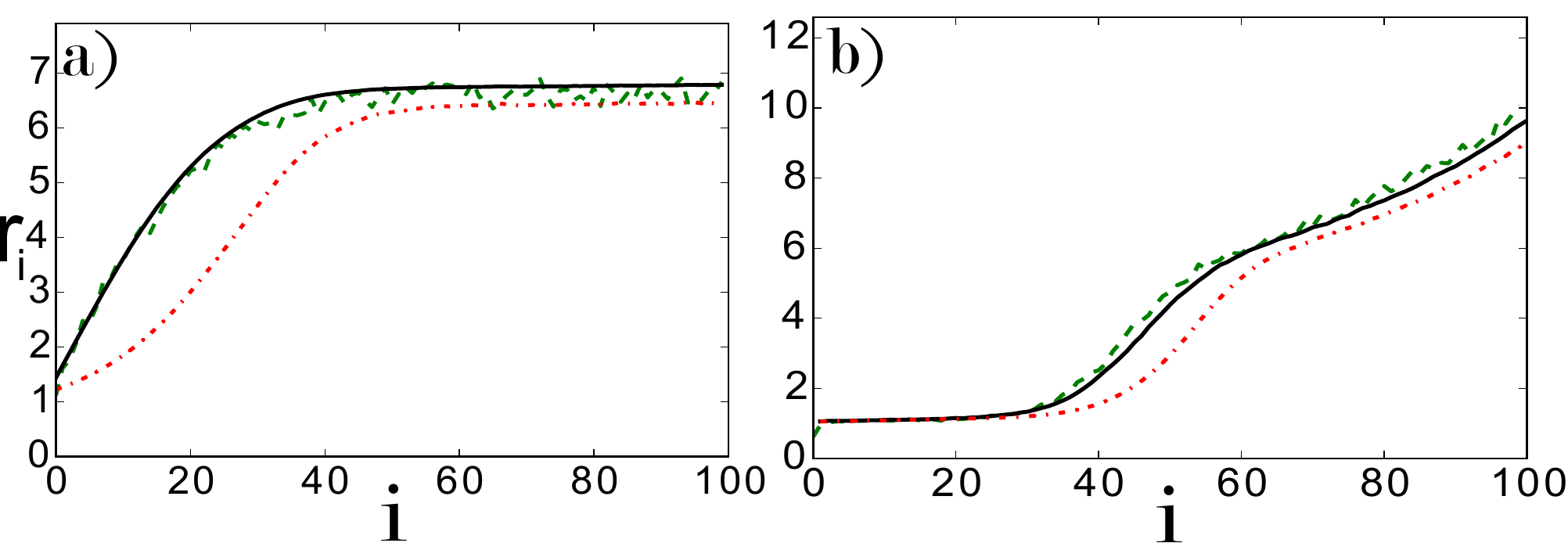}
\caption{Residence time of a probe particle in presence of system particles undergoing TASEP + MILK. (a) Symmetrically modified kinetic rates $\delta = \gamma =0$, $\alpha = \beta = 0.15$, $\Omega = 0.1$, $N=100$. (b) Asymmetrically modified kinetic rates $\delta = 1 + \psi$ and $\gamma = 1-\psi$ with $\psi = 0.2$, $\alpha = 0.05$, $\beta = 0.15$, $\Omega = 0.3$, $N=100$. Green dashed lines: residence time of the probe particle from simulations. Thick solid lines: Our theoretical prediction. Red dash-dotted lines: Mean-field prediction of the on-site residence time.}
\label{MILKprofiles}
\end{figure}

The solutions for the density profile are only known in specific regimes. We therefore compute the residence times in each of these regimes separately. We first consider the the parameter regime $\delta = \gamma$. In this case both attachment and detachment rates are symmetrically modified such that both the attachment and detachment rates are either scaled up or scaled down. The solution to the eq.~\eqref{diffMILK} derived in Ref.~\cite{vuijk2015driven} is reproduced in the Appendix. Using eqs.~\eqref{rholowcase1} and \eqref{rhohighcase1} together with eqs.~\eqref{p} and ~\eqref{rhofromp} we obtain the density profile for a given system size $N$. We use numerical integration to obtain the on-site residence time from eqs.~\eqref{rnew} and ~\eqref{lambda} which is shown in Fig.~\ref{MILKprofiles}.

Another interesting parameter regime is $\delta = 1 + \psi, \gamma = 1-\psi$. In this case mutual interactions modify the kinetic rates in an asymmetric fashion such that one is scaled up and the other is scaled down. Assuming that $\psi \ll 1$ one can obtain closed form expression for the solution to eq.~\eqref{diffMILK} as shown in Ref.~\cite{vuijk2015driven}. The expressions are reproduced in the Appendix (eqs.~\eqref{eq:rho_a case 2} and \eqref{eq:rho_b case 2}). Performing similar calculations as in the case of symmetrically modified kinetic rates, we obtain on-site residence time as shown in Fig.~\ref{MILKprofiles}.

As can be seen in Fig.~\ref{MILKprofiles}, our theoretical predictions are in excellent agreement with the numerically obtained on-site residence time. We believe that using the concept of probe particle, our theoretical approach can describe the residence time in presence of interactions other than the MILK.

\section{Transition to mean-field theory in LD-HD phase}\label{limitations}
The approach for calculating on-site residence times presented in this article is valid for any two-phase system that obeys the two conditions described in section \ref{ourtheory}. To explore a system for which this approach does not work, we introduce a \textit{ghost particle} to our system which we give the following properties:
\begin{itemize}
\item The particle is first placed on the first site.
\item The particle moves to the next site if this site is unoccupied.
\item Other system particles are not influenced by this particle. The system particles can thus hop onto a site if it is occupied by the ghost particle, as long as it is not occupied by another system particle.
\item When the ghost particle reaches site $N$, it has a probability of $\beta$ of detaching.
\end{itemize}
Furthermore, we will give this particle a probability to \textit{pause}, a concept introduced in Ref.~\cite{wang2014minimal}. When a particle is paused, it will not be able to hop. As long as the particle is not paused, it will move according to the rules above. We give the particle a probability of $f$ to be in the active state on a single timestep and thus a probability of $1-f$ to be in the paused state. We keep track of the on-site residence times of this particle over all timesteps that it is in the active phase.

\begin{figure}
\includegraphics[width = \columnwidth]{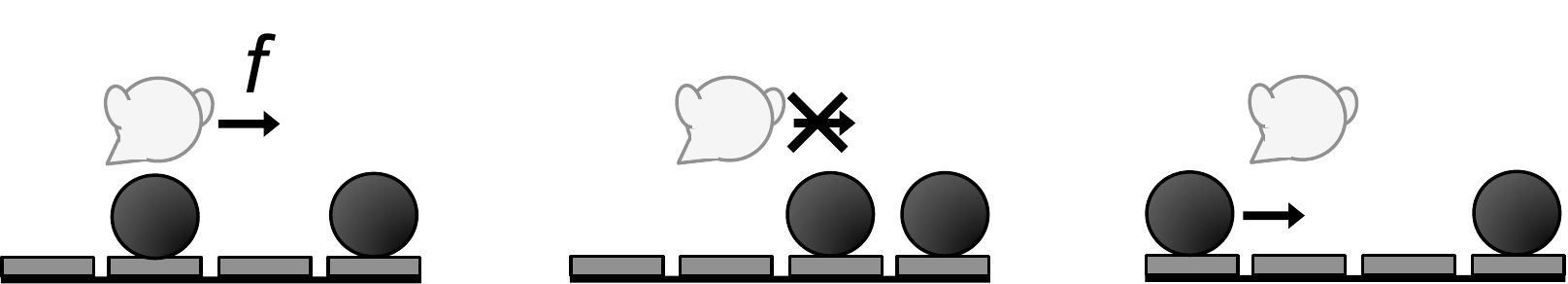}
\caption{A graphic representation of the ghost particle dynamics. The ghost particle is first placed on the first site. When in active state, the ghost particle moves exclusively to the right only if the next site is unoccupied. In the inactive state, the ghost particle always remains at the same site. The probability of being in active state is denoted by $f$. On reaching site $N$, the ghost particle detaches with a rate of $\beta$. Effectively, the ghost particle follows TASEP in active state. Other system particles do not interact with the ghost particle. A site can be occupied by a system particle together with the ghost particle.}
\end{figure}

Due to these movement rules for our ghost particle, it is now possible for the flag described in section \ref{flaglkrs} to pass under the second-class particle while it is waiting to hop. The second assumption of our approach, that a particle only switches between phases while making a hop, will thus not be satisfied if $f<1$. We thus expect the prediction of our theory for the crossing time of the ghost particle to be inaccurate in this case. If the probability to be in the active state $f$ is low enough, the average time intervals between active states of the paused particle will be larger than the time interval over which we observe time correlations. We then expect for approximation \eqref{MFexpression} to be valid again, and thus for the mean-field expression for the on-site residence times to be accurate. In the intermediate regime, we expect both approaches to be inaccurate, and we expect the on-site residence times to be somewhere between the mean-field value and the value predicted by eq.~\eqref{rnew}.

\begin{figure}
\includegraphics[width = \columnwidth]{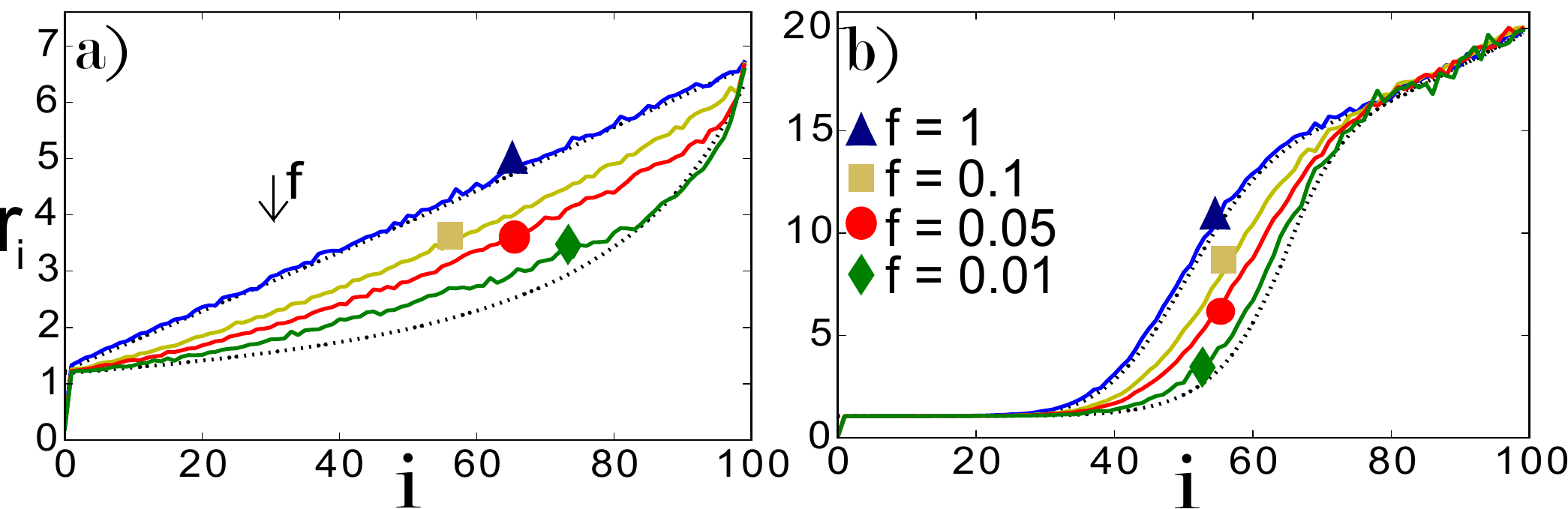}
\caption{Plot of the on-site residence time of the ghost particle for different values of ghost frequencies $f$. The dashed line and the dotted line indicate the on-site residence times predicted by eq.~\eqref{rnew} and by mean-field theory respectively. The dashed line coincides with the residence time of the ghost particle for $f = 1$.}
\label{ghostplots}
\end{figure}

We ran simulations for the on-site residence times of this ghost particle for various values of $f$, for two LD-HD profiles. The results are shown in Fig.~\ref{ghostplots}. As can be seen in this figure, the on-site residence times for the ghost particle indeed make a gradual transition from the prediction of eq.~\eqref{rnew} to the mean-field prediction for decreasing $f$.

\section{Discussion and conclusion}\label{conclusions}
In this paper, we study the average residence time of a particle on a given site in a 1D driven diffusive system with open boundaries. The particle performs totally asymmetric motion with hard core repulsion. In a mean-field scenario, ignorning spatial and temporal correlations, we obtain the on-site residence time as inverse of the average hopping rate $(1-\rho_i(1-\omega_D))$, where $\rho_i$ is the steady-state density of particles at site $i$ and $\omega_D$ is the single-site detachment probability. Using Monte Carlo simulations we show that the mean-field significantly understimates the residence time in the neighborhood of shock interface. The shock interface is characterized by a steep transition in the density profile of particles in the steady state. In the thermodynamic limit, the transition is infinitely steep, reminiscent of a first order phase transition with two-phase coexistence. The underlying reason for failure of the mean-field estimate for residence time is the neglect of time-correlations in the local density of particles. The temporal correlations in the density are especially long lived near the shock interface with the local correlation time much larger than the mean-field on-site residence time. 

The shock interface separating the low and high density phases is not localized on a given site. In pure TASEP, the interface is delocalized over the entire lattice for any system size. On adding LK, the interface becomes localized within a region that scales with $\sqrt{N}$ where $N$ is the system size. For any finite size lattice, the interface can be considered as performing random walk in a confined potential. The mean-field residence time does not take into account the fact that the density at a local site fluctuates on time-scale longer than the mean-field estimate.  One can obtain the steady state density profile at a given site by averaging over time scales longer than the correlation time. However, the presence of long-lived correlations leads to an underestimation of the residence time using mean-field. We provide a non-mean-field expression for the on-site residence time which takes the movement of the shock interface into account. Our description requires calculation of a single parameter, referred to as the site-dependent, average phase constant. Our derivation of a non mean-field equation for the on-site residence time relies on two assumptions: (1) that each site can at each timestep be determined to be either in the low or the high density phase, (2) that a particle crossing the lattice switches phases only while making a hop. These two assumptions are validated by considering Flag theory for the movement of the shock. In previous studies, Flag theory has been successfully applied to pure TASEP to track the shock location. Here we extend the Flag theory to TASEP + LK and determine the flag movement rules. These flag movement rules validate the assumptions of our non mean-field derivation of on-site residence times.  

We obtain the average phase constant making use of previous results for the average density profile from domain wall dynamics and compare the analytical residence time with the numerical simulations. We show that our analytical predictions are highly accurate in describing the residence time in TASEP as well as TASEP+LK.

The total residence time on a pure TASEP lattice can be obtained as the sum of the on-site residence times. In TASEP+LK, the possibility of a particle detaching before reaching the $\beta$ boundary also needs to be taken into account. Doing this yields a weighted sum of the on-site residence times as an expression for the total residence time. We demonstrate that the total residence time asymptotically scales linearly with the system size.

Our approach to calculating residence time can be extended to TASEP systems with further modified dynamics by considering a test particle. This is a particle that does not detach and will thus cross the entire lattice. The on-site residence time for this particle is obtainable provided $\rho_{\mathrm{HD}}$, $\rho_{\mathrm{LD}}$ and $\rho_(x)$ are known. We run simulations for the on-site residence time of a test particle in the TASEP+MILK system and find good agreement with our predictions.

In order to demonstrate how mean-field theory can be recovered by violating the second assumption of our approach, we consider a ghost particle on the lattice which performs pure TASEP. The ghost particle has a hopping rate that is lower than that of the system particles. The ghost particle, somewhat similar to a second class particle, is invisible to other particles but is itself subjected to hard core repulsion. We demonstrate that when the hopping rate of the ghost particle is sufficiently low, it can experience phase-change without performing a hop. The phase change occurs due to the shock interface moving past the ghost particle while it is paused. In the case of vanishing hopping rate, we find that the residence time of the ghost particle is accurately captured by the mean-field estimate. 

Our focus in this study is the residence time on a lattice of given size with given boundary conditions. It will be interesting to explore the same in a network setting, viewed as collection of such 1D systems. The boundary conditions in a network will be specified only on the outermost boundary with all the internal nodes exhibiting fluctuations in the injection and extraction rates. Such a study could be linked to existing work considering TASEP on networks \cite{Neri2011totally} and TASEP+LK on networks \cite{Neri2013exclusion}. Another interesting study would be to apply the concept of probe particle to systems with modified interactions. Finally, it is a challenging and interesting problem to accurately describe the residence time of ghost particle for any given hopping rate.

\appendix 
\section{Solutions for $\rho_{\mathrm{HD}}$ and $\rho_{\mathrm{LD}}$ for TASEP+MILK}

\subsection{The case $\delta = \gamma$}
In the case we set $\delta = \gamma = 1+\eta$, the two solutions to \eqref{diffMILK} that meet the boundary conditions as derived in Ref.~\cite{vuijk2015driven} are

\begin{equation} \label{rholowcase1}
\rho_{\mathrm{LD}}= \frac{\alpha + (1+ \eta \alpha) \Omega x}{1-(1+\eta \alpha)\eta \Omega x}
\end{equation}
\begin{equation}\label{rhohighcase1}
\rho_{\mathrm{HD}}=\frac{1 -\beta + (\eta(1 - \beta) + 1) \Omega(x-1)}{1-(\eta(1 - \beta) +1) \eta \Omega (x-1)}.
\end{equation}

A shock forms in the case that $x_-^I > x_+^I$, where $x_-^I$ is the value for $x$ at which $\rho_{\mathrm{LD}}$ crosses the isotherm of $\rho_I = \frac12$, and $x_+^I$   is the value for $x$ at which $\rho_{\mathrm{HD}}$ crosses the isotherm.

\subsection{The case $\delta = 1 + \psi, \gamma = 1-\psi$}
The two solutions to \eqref{diffMILK} that meet the boundary conditions as derived in Ref.~\cite{vuijk2015driven} are
\begin{equation}\label{eq:rho_a case 2}
\rho_{\mathrm{LD}}=\frac{\psi}{2(1-\psi)} \left( W_{-1} \left[-y(x) \right] + 1 \right) + \frac{1}{2} \text{\ \ for $\alpha<1/2$},
\end{equation}
\begin{equation}\label{eq:rho_b case 2}
\\ \rho_{\mathrm{HD}} = \begin{cases}
\frac{\psi}{2(1-\psi)} \left( W_0 \left[y(x) \right] + 1 \right) + \frac{1}{2} & \text{for $1-\beta \geq \rho_I$},\\
\frac{\psi}{2(1-\psi)} \left( W_0 \left[-y(x) \right] + 1 \right) + \frac{1}{2} & \text{for $\frac{1}{2} \leq 1-\beta \leq \rho_I$},\\
\end{cases}
\end{equation}
where $\rho_{\mathrm{LD}}$ obeys the left and $\rho_{\mathrm{HD}}$ the right boundary condition. $W[y]$ is the Lambert-W function. $y(x)$ Is given by
\begin{multline}\label{y(x)}
y(x)=\left |\frac{1-\psi}{\psi} (2\rho_0-1) -1\right| \\exp \left[2\Omega \frac{(1-\psi)^2}{\psi} (x-x_0)+\frac{1-\psi}{\psi} (2\rho_0-1) -1\right],
\end{multline}
with $\rho_0=\alpha$, $x_0=0$ for $\rho_{\mathrm{LD}}$ and $\rho_0=1-\beta$, $x_0=1$ for $\rho_{\mathrm{HD}}$. The constant solution $\rho_{MC}=\frac{1}{2(1-\psi)}$ is the equivalent of the Langmuir isotherm in the case without MI. 
The solution $\rho_{\mathrm{LD}}$ is stable only for $\alpha<1/2$, and $\rho_{\mathrm{HD}}$ for $\beta \leq 1/2$.

A shock will again form in the case that $x_{\mathrm{LD}}^I > x_{\mathrm{HD}}^I$, where $x_{\mathrm{LD}}^I$ is the value for $x$ at which $\rho_{\mathrm{LD}}$ crosses the isotherm and $x_{\mathrm{HD}}^I$ is the value for $x$ at which $\rho_{\mathrm{HD}}$ crosses the isotherm.

\begin{widetext}
\section{Linearity of R for the LD-HD phase in the large N limit}\label{linearR}
In the large $N$ limit, we can approximate the shock to be completely localized at $x_s$. The average density will then be equal to $\rho_{\mathrm{LD}}$ from 0 to $x_s$, and equal to $\rho_{\mathrm{HD}}$ from $x_s$ to 1. We can thus approximate $R$ as
\begin{align}\label{RlargeN}
R = &r_1 + \sum_{n=2}^{n_s} r_n^{\mathrm{LD}} \prod_{m=2}^{n} (1-\omega_D r_m^{\mathrm{LD}}) + \sum_{n=n_s+1}^N r_n^{\mathrm{HD}} \left[ \prod_{m=2}^{n_s} (1-\omega_D r_m^{\mathrm{LD}}) \right] \left[\prod_{m=n_s+1}^{n} (1-\omega_D r_m^{\mathrm{HD}}) \right]. 
\end{align}
Here, $r_n^{\mathrm{LD}}$ is the on-site residence time corresponding to $\rho_{\mathrm{LD},n+1}$ and $r_n^{\mathrm{HD}}$ the on-site residence time corresponding to $\rho_{\mathrm{HD},n+1}$. In order to evaluate this expression, we will make a few approximations. The product terms we will rewrite as follows:
\begin{align}\label{productR}
\prod_{m=2}^{n} (1-\omega_D r_m^{\mathrm{LD}}) &= 1 - \omega_D \sum_{m=2}^{n} r_m^{\mathrm{LD}} + \omega_D^2 \left[ \frac{1}{2} \left(\sum_{m=2}^{n} r_m^{\mathrm{LD}} \right)^2 - \sum_{m=2}^{n} (r_m^{\mathrm{LD}})^2\right] \nonumber \\
&- \omega_D^3 \left[ \frac{1}{3!} \left(\sum_{m=2}^{n} r_m^{\mathrm{LD}} \right)^3 - \sum_{m=2}^{n} (r_m^{LD})^2 \left( \sum_{m=2}^{n} (r_m^{\mathrm{LD}})\right) \right] + ... \nonumber \\
&= \underbrace{\sum_{j=0}^{n} (-1)^j \omega_D^j \frac{1}{j!} \left(\sum_{m=2}^{n} (r_m^{\mathrm{LD}})\right)^j}_\text{\raisebox{.5pt}{\textcircled{\raisebox{-.9pt} {1}}}} - \underbrace{\sum_{j=2}^{n} (-1)^j \omega_D^j \frac{1}{(j-2)!} \sum_{m=2}^{n}((r_m^{\mathrm{LD}})^2)\left(\sum_{m=2}^{n}(r_m^{\mathrm{LD}})\right)^{j-2}}_\text{\raisebox{.5pt}{\textcircled{\raisebox{-.9pt} {2}}}} .
\end{align}
And similarly for the product over $1-\omega_D r_m^{\mathrm{HD}}$. The second term in each of the square brackets ensures that the quantity in square brackets contains only products of terms for which $m \neq m'$. The factors $\frac{1}{j!}$ and $\frac{1}{(j-2)!}$ ensure there is no double counting. 

We will now investigate the components of eq.~\eqref{productR} separately. The sum over $m$ in $\raisebox{.5pt}{\textcircled{\raisebox{-.9pt} {1}}}$ we can approximate as

\begin{align}
\sum_{m=2}^{n} (r_m^{\mathrm{LD}}) &= \sum_{m=2}^{n} \frac{1}{1-(\frac{m}{N}\Omega + \alpha)} \nonumber \\
&\approx \int_0^n \frac{1}{1-(\frac{m}{N}\Omega + \alpha)} \mathrm{d}m = - \frac{N}{\Omega} \ln \left(\frac{1-(\Omega \frac{n}{N} + \alpha)}{1 - \alpha}\right).
\end{align}
Furthermore, the first sum over $m$ in $\raisebox{.5pt}{\textcircled{\raisebox{-.9pt} {2}}}$ can be approximated as
\begin{align}
\sum_{m=2}^{n}((r_m^{\mathrm{LD}})^2) \approx \int_0^n \frac{1}{(1-(\frac{m}{N}\Omega + \alpha))^2} \mathrm{d}m = \frac{N}{\Omega}  \frac{1}{(1-(\frac{m}{N}\Omega + \alpha))}.
\end{align}
We will now look at how the different terms in eq.~\eqref{productR} scale with $N$. As $\omega_D = \frac{\Omega}{N}$ in the mesoscopic scaling, we see that $\omega_D^j \times \raisebox{.5pt}{\textcircled{\raisebox{-.9pt} {1}}}$ scales as $N^0$, while $\omega_D^j \times \raisebox{.5pt}{\textcircled{\raisebox{-.9pt} {2}}}$ scales as $N^{-1}$. Thus, in the large $N$ limit we can discard $\raisebox{.5pt}{\textcircled{\raisebox{-.9pt} {2}}}$ in eq.~\eqref{productR}.

We are then left with the following expression for eq.~\eqref{productR}:
\begin{align}\label{prodLD}
\prod_{m=2}^{n} (1-\omega_D r_m^{LD}) \approx \sum_{j=1}^{n} (-1)^j \frac{1}{j!} \left[ -\ln \left(\frac{1-(\Omega \frac{n}{N} + \alpha)}{1 - \alpha}\right)\right]^j \approx \frac{1-(\Omega \frac{n}{N} + \alpha)}{1 - \alpha}.
\end{align}
In the second step we used the identity $\lim_{n \rightarrow \infty} \sum_{j=1}^{n} \frac{1}{j!} x^j = e^x$.

A similar approximation can be made for $\prod_{m=2}^{n} (1-\omega_D r_m^{HD}) $, the result is
\begin{align}\label{prodHD}
\prod_{m=n_s+1}^{n} (1-\omega_D r_m^{HD})  \approx \frac{\Omega(1- \frac{n}{N}) + \beta}{\Omega(1- \frac{n_s}{N}) + \beta}.
\end{align}

We will now insert the approximations for the product terms into eq.~\eqref{RlargeN} and convert the sum over $n$ to an integral. Evaluating this integral yields the final result

\begin{align}\label{Rlinear}
R = \frac{n_s}{1-\alpha} + \left(1-\frac{\Omega \frac{n_s}{N}}{1-\alpha} \right) \frac{(N-n_s)}{\Omega (1-\frac{n_s}{N}) + \beta}
\end{align}
which is indeed predominantly linear in the number of sites $N$ (note that in the mesoscopic scaling, $n_s$ is a constant fraction of $N$). An evaluation of eq.~\eqref{Rlinear} for explicit choices for the model parameters is plotted together with the results from Monte Carlo simulation in Fig.~\ref{allrs}.

\end{widetext}

\bibliographystyle{apsrev}
\bibliography{Bibliography}

\begin{thebibliography}{47}
\expandafter\ifx\csname natexlab\endcsname\relax\def\natexlab#1{#1}\fi
\expandafter\ifx\csname bibnamefont\endcsname\relax
  \def\bibnamefont#1{#1}\fi
\expandafter\ifx\csname bibfnamefont\endcsname\relax
  \def\bibfnamefont#1{#1}\fi
\expandafter\ifx\csname citenamefont\endcsname\relax
  \def\citenamefont#1{#1}\fi
\expandafter\ifx\csname url\endcsname\relax
  \def\url#1{\texttt{#1}}\fi
\expandafter\ifx\csname urlprefix\endcsname\relax\def\urlprefix{URL }\fi
\providecommand{\bibinfo}[2]{#2}
\providecommand{\eprint}[2][]{\url{#2}}

\bibitem[{\citenamefont{Krug}(1991)}]{krug1991boundary}
\bibinfo{author}{\bibfnamefont{J.}~\bibnamefont{Krug}}, \bibinfo{journal}{Phys.
  Rev. Lett.} \textbf{\bibinfo{volume}{67}}, \bibinfo{pages}{1882}
  (\bibinfo{year}{1991}).

\bibitem[{\citenamefont{Derrida et~al.}(1993)\citenamefont{Derrida, Evans,
  Hakim, and Pasquier}}]{derrida1993exact}
\bibinfo{author}{\bibfnamefont{B.}~\bibnamefont{Derrida}},
  \bibinfo{author}{\bibfnamefont{M.~R.} \bibnamefont{Evans}},
  \bibinfo{author}{\bibfnamefont{V.}~\bibnamefont{Hakim}}, \bibnamefont{and}
  \bibinfo{author}{\bibfnamefont{V.}~\bibnamefont{Pasquier}},
  \bibinfo{journal}{Journal of Physics A: Mathematical and General}
  \textbf{\bibinfo{volume}{26}}, \bibinfo{pages}{1493} (\bibinfo{year}{1993}).

\bibitem[{\citenamefont{Evans et~al.}(1995)\citenamefont{Evans, Foster,
  Godreche, and Mukamel}}]{evans1995spontaneous}
\bibinfo{author}{\bibfnamefont{M.}~\bibnamefont{Evans}},
  \bibinfo{author}{\bibfnamefont{D.}~\bibnamefont{Foster}},
  \bibinfo{author}{\bibfnamefont{C.}~\bibnamefont{Godreche}}, \bibnamefont{and}
  \bibinfo{author}{\bibfnamefont{D.}~\bibnamefont{Mukamel}},
  \bibinfo{journal}{Physical review letters} \textbf{\bibinfo{volume}{74}},
  \bibinfo{pages}{208} (\bibinfo{year}{1995}).

\bibitem[{\citenamefont{Domb et~al.}(1995)\citenamefont{Domb, Zia, Schmittmann,
  and Lebowitz}}]{domb1995statistical}
\bibinfo{author}{\bibfnamefont{C.}~\bibnamefont{Domb}},
  \bibinfo{author}{\bibfnamefont{R.~K.} \bibnamefont{Zia}},
  \bibinfo{author}{\bibfnamefont{B.}~\bibnamefont{Schmittmann}},
  \bibnamefont{and} \bibinfo{author}{\bibfnamefont{J.~L.}
  \bibnamefont{Lebowitz}}, \emph{\bibinfo{title}{Statistical mechanics of
  driven diffusive systems}}, vol.~\bibinfo{volume}{17}
  (\bibinfo{publisher}{Academic Press}, \bibinfo{year}{1995}).

\bibitem[{\citenamefont{Evans et~al.}(1998)\citenamefont{Evans, Kafri,
  Koduvely, and Mukamel}}]{evans1998phase}
\bibinfo{author}{\bibfnamefont{M.}~\bibnamefont{Evans}},
  \bibinfo{author}{\bibfnamefont{Y.}~\bibnamefont{Kafri}},
  \bibinfo{author}{\bibfnamefont{H.}~\bibnamefont{Koduvely}}, \bibnamefont{and}
  \bibinfo{author}{\bibfnamefont{D.}~\bibnamefont{Mukamel}},
  \bibinfo{journal}{Physical review letters} \textbf{\bibinfo{volume}{80}},
  \bibinfo{pages}{425} (\bibinfo{year}{1998}).

\bibitem[{\citenamefont{Oloan et~al.}(1998)\citenamefont{Oloan, Evans, and
  Cates}}]{o1998jamming}
\bibinfo{author}{\bibfnamefont{O.}~\bibnamefont{Oloan}},
  \bibinfo{author}{\bibfnamefont{M.}~\bibnamefont{Evans}}, \bibnamefont{and}
  \bibinfo{author}{\bibfnamefont{M.}~\bibnamefont{Cates}},
  \bibinfo{journal}{Physical Review E} \textbf{\bibinfo{volume}{58}},
  \bibinfo{pages}{1404} (\bibinfo{year}{1998}).

\bibitem[{\citenamefont{Schmittmann and Zia}(1998)}]{schmittmann1998driven}
\bibinfo{author}{\bibfnamefont{B.}~\bibnamefont{Schmittmann}} \bibnamefont{and}
  \bibinfo{author}{\bibfnamefont{R.}~\bibnamefont{Zia}},
  \bibinfo{journal}{Physics reports} \textbf{\bibinfo{volume}{301}},
  \bibinfo{pages}{45} (\bibinfo{year}{1998}).

\bibitem[{\citenamefont{Helbing}(2001)}]{helbing2001traffic}
\bibinfo{author}{\bibfnamefont{D.}~\bibnamefont{Helbing}},
  \bibinfo{journal}{Rev. Mod. Phys.} \textbf{\bibinfo{volume}{73}},
  \bibinfo{pages}{1067} (\bibinfo{year}{2001}).

\bibitem[{\citenamefont{Parmeggiani et~al.}(2003)\citenamefont{Parmeggiani,
  Franosch, and Frey}}]{parmeggiani2003phase}
\bibinfo{author}{\bibfnamefont{A.}~\bibnamefont{Parmeggiani}},
  \bibinfo{author}{\bibfnamefont{T.}~\bibnamefont{Franosch}}, \bibnamefont{and}
  \bibinfo{author}{\bibfnamefont{E.}~\bibnamefont{Frey}},
  \bibinfo{journal}{Phys. Rev. Lett.} \textbf{\bibinfo{volume}{90}},
  \bibinfo{pages}{086601} (\bibinfo{year}{2003}).

\bibitem[{\citenamefont{Parmeggiani et~al.}(2004)\citenamefont{Parmeggiani,
  Franosch, and Frey}}]{parmeggiani2004totally}
\bibinfo{author}{\bibfnamefont{A.}~\bibnamefont{Parmeggiani}},
  \bibinfo{author}{\bibfnamefont{T.}~\bibnamefont{Franosch}}, \bibnamefont{and}
  \bibinfo{author}{\bibfnamefont{E.}~\bibnamefont{Frey}},
  \bibinfo{journal}{Phys. Rev. E} \textbf{\bibinfo{volume}{70}},
  \bibinfo{pages}{046101} (\bibinfo{year}{2004}).

\bibitem[{\citenamefont{Kwak et~al.}(2004)\citenamefont{Kwak, Landau, and
  Schmittmann}}]{kwak2004driven}
\bibinfo{author}{\bibfnamefont{W.}~\bibnamefont{Kwak}},
  \bibinfo{author}{\bibfnamefont{D.}~\bibnamefont{Landau}}, \bibnamefont{and}
  \bibinfo{author}{\bibfnamefont{B.}~\bibnamefont{Schmittmann}},
  \bibinfo{journal}{Physical Review E} \textbf{\bibinfo{volume}{69}},
  \bibinfo{pages}{066134} (\bibinfo{year}{2004}).

\bibitem[{\citenamefont{Kafri et~al.}(2002)\citenamefont{Kafri, Levine,
  Mukamel, Sch{\"u}tz, and T{\"o}r{\"o}k}}]{kafri2002criterion}
\bibinfo{author}{\bibfnamefont{Y.}~\bibnamefont{Kafri}},
  \bibinfo{author}{\bibfnamefont{E.}~\bibnamefont{Levine}},
  \bibinfo{author}{\bibfnamefont{D.}~\bibnamefont{Mukamel}},
  \bibinfo{author}{\bibfnamefont{G.}~\bibnamefont{Sch{\"u}tz}},
  \bibnamefont{and}
  \bibinfo{author}{\bibfnamefont{J.}~\bibnamefont{T{\"o}r{\"o}k}},
  \bibinfo{journal}{Physical review letters} \textbf{\bibinfo{volume}{89}},
  \bibinfo{pages}{035702} (\bibinfo{year}{2002}).

\bibitem[{\citenamefont{Appert-Rolland
  et~al.}(2015)\citenamefont{Appert-Rolland, Ebbinghaus, and
  Santen}}]{appert2015intracellular}
\bibinfo{author}{\bibfnamefont{C.}~\bibnamefont{Appert-Rolland}},
  \bibinfo{author}{\bibfnamefont{M.}~\bibnamefont{Ebbinghaus}},
  \bibnamefont{and} \bibinfo{author}{\bibfnamefont{L.}~\bibnamefont{Santen}},
  \bibinfo{journal}{Physics Reports} \textbf{\bibinfo{volume}{593}},
  \bibinfo{pages}{1} (\bibinfo{year}{2015}).

\bibitem[{\citenamefont{Shaw et~al.}(2003)\citenamefont{Shaw, Zia, and
  Lee}}]{shaw2003totally}
\bibinfo{author}{\bibfnamefont{L.~B.} \bibnamefont{Shaw}},
  \bibinfo{author}{\bibfnamefont{R.}~\bibnamefont{Zia}}, \bibnamefont{and}
  \bibinfo{author}{\bibfnamefont{K.~H.} \bibnamefont{Lee}},
  \bibinfo{journal}{Physical Review E} \textbf{\bibinfo{volume}{68}},
  \bibinfo{pages}{021910} (\bibinfo{year}{2003}).

\bibitem[{\citenamefont{Ciandrini et~al.}(2010)\citenamefont{Ciandrini,
  Stansfield, and Romano}}]{ciandrini2010role}
\bibinfo{author}{\bibfnamefont{L.}~\bibnamefont{Ciandrini}},
  \bibinfo{author}{\bibfnamefont{I.}~\bibnamefont{Stansfield}},
  \bibnamefont{and} \bibinfo{author}{\bibfnamefont{M.~C.}
  \bibnamefont{Romano}}, \bibinfo{journal}{Physical Review E}
  \textbf{\bibinfo{volume}{81}}, \bibinfo{pages}{051904}
  (\bibinfo{year}{2010}).

\bibitem[{\citenamefont{Basu and Chowdhury}(2007)}]{basu2007traffic}
\bibinfo{author}{\bibfnamefont{A.}~\bibnamefont{Basu}} \bibnamefont{and}
  \bibinfo{author}{\bibfnamefont{D.}~\bibnamefont{Chowdhury}},
  \bibinfo{journal}{Physical Review E} \textbf{\bibinfo{volume}{75}},
  \bibinfo{pages}{021902} (\bibinfo{year}{2007}).

\bibitem[{\citenamefont{Reuveni et~al.}(2011)\citenamefont{Reuveni, Meilijson,
  Kupiec, Ruppin, and Tuller}}]{reuveni2011genome}
\bibinfo{author}{\bibfnamefont{S.}~\bibnamefont{Reuveni}},
  \bibinfo{author}{\bibfnamefont{I.}~\bibnamefont{Meilijson}},
  \bibinfo{author}{\bibfnamefont{M.}~\bibnamefont{Kupiec}},
  \bibinfo{author}{\bibfnamefont{E.}~\bibnamefont{Ruppin}}, \bibnamefont{and}
  \bibinfo{author}{\bibfnamefont{T.}~\bibnamefont{Tuller}},
  \bibinfo{journal}{PLoS Comput Biol} \textbf{\bibinfo{volume}{7}},
  \bibinfo{pages}{e1002127} (\bibinfo{year}{2011}).

\bibitem[{\citenamefont{Heinrich and
  Rapoport}(1980)}]{heinrich1980mathematical}
\bibinfo{author}{\bibfnamefont{R.}~\bibnamefont{Heinrich}} \bibnamefont{and}
  \bibinfo{author}{\bibfnamefont{T.~A.} \bibnamefont{Rapoport}},
  \bibinfo{journal}{Journal of theoretical biology}
  \textbf{\bibinfo{volume}{86}}, \bibinfo{pages}{279} (\bibinfo{year}{1980}).

\bibitem[{\citenamefont{Chowdhury et~al.}(2005)\citenamefont{Chowdhury,
  Schadschneider, and Nishinari}}]{chowdhury2005physics}
\bibinfo{author}{\bibfnamefont{D.}~\bibnamefont{Chowdhury}},
  \bibinfo{author}{\bibfnamefont{A.}~\bibnamefont{Schadschneider}},
  \bibnamefont{and}
  \bibinfo{author}{\bibfnamefont{K.}~\bibnamefont{Nishinari}},
  \bibinfo{journal}{Physics of Life reviews} \textbf{\bibinfo{volume}{2}},
  \bibinfo{pages}{318} (\bibinfo{year}{2005}).

\bibitem[{\citenamefont{Kural et~al.}(2007)\citenamefont{Kural, Serpinskaya,
  Chou, Goldman, Gelfand, and Selvin}}]{kural2007tracking}
\bibinfo{author}{\bibfnamefont{C.}~\bibnamefont{Kural}},
  \bibinfo{author}{\bibfnamefont{A.~S.} \bibnamefont{Serpinskaya}},
  \bibinfo{author}{\bibfnamefont{Y.-H.} \bibnamefont{Chou}},
  \bibinfo{author}{\bibfnamefont{R.~D.} \bibnamefont{Goldman}},
  \bibinfo{author}{\bibfnamefont{V.~I.} \bibnamefont{Gelfand}},
  \bibnamefont{and} \bibinfo{author}{\bibfnamefont{P.~R.}
  \bibnamefont{Selvin}}, \bibinfo{journal}{Proceedings of the National Academy
  of Sciences} \textbf{\bibinfo{volume}{104}}, \bibinfo{pages}{5378}
  (\bibinfo{year}{2007}).

\bibitem[{\citenamefont{Churchman et~al.}(2005)\citenamefont{Churchman,
  {\"O}kten, Rock, Dawson, and Spudich}}]{churchman2005single}
\bibinfo{author}{\bibfnamefont{L.~S.} \bibnamefont{Churchman}},
  \bibinfo{author}{\bibfnamefont{Z.}~\bibnamefont{{\"O}kten}},
  \bibinfo{author}{\bibfnamefont{R.~S.} \bibnamefont{Rock}},
  \bibinfo{author}{\bibfnamefont{J.~F.} \bibnamefont{Dawson}},
  \bibnamefont{and} \bibinfo{author}{\bibfnamefont{J.~A.}
  \bibnamefont{Spudich}}, \bibinfo{journal}{Proceedings of the National Academy
  of Sciences of the United States of America} \textbf{\bibinfo{volume}{102}},
  \bibinfo{pages}{1419} (\bibinfo{year}{2005}).

\bibitem[{\citenamefont{Yildiz and Selvin}(2005)}]{yildiz2005fluorescence}
\bibinfo{author}{\bibfnamefont{A.}~\bibnamefont{Yildiz}} \bibnamefont{and}
  \bibinfo{author}{\bibfnamefont{P.~R.} \bibnamefont{Selvin}},
  \bibinfo{journal}{Accounts of chemical research}
  \textbf{\bibinfo{volume}{38}}, \bibinfo{pages}{574} (\bibinfo{year}{2005}).

\bibitem[{\citenamefont{Courty et~al.}(2006)\citenamefont{Courty, Luccardini,
  Bellaiche, Cappello, and Dahan}}]{courty2006tracking}
\bibinfo{author}{\bibfnamefont{S.}~\bibnamefont{Courty}},
  \bibinfo{author}{\bibfnamefont{C.}~\bibnamefont{Luccardini}},
  \bibinfo{author}{\bibfnamefont{Y.}~\bibnamefont{Bellaiche}},
  \bibinfo{author}{\bibfnamefont{G.}~\bibnamefont{Cappello}}, \bibnamefont{and}
  \bibinfo{author}{\bibfnamefont{M.}~\bibnamefont{Dahan}},
  \bibinfo{journal}{Nano letters} \textbf{\bibinfo{volume}{6}},
  \bibinfo{pages}{1491} (\bibinfo{year}{2006}).

\bibitem[{\citenamefont{Harada et~al.}(2001)\citenamefont{Harada, Ohara,
  Takatsuki, Itoh, Shimamoto, and Kinosita}}]{harada2001direct}
\bibinfo{author}{\bibfnamefont{Y.}~\bibnamefont{Harada}},
  \bibinfo{author}{\bibfnamefont{O.}~\bibnamefont{Ohara}},
  \bibinfo{author}{\bibfnamefont{A.}~\bibnamefont{Takatsuki}},
  \bibinfo{author}{\bibfnamefont{H.}~\bibnamefont{Itoh}},
  \bibinfo{author}{\bibfnamefont{N.}~\bibnamefont{Shimamoto}},
  \bibnamefont{and} \bibinfo{author}{\bibfnamefont{K.}~\bibnamefont{Kinosita}},
  \bibinfo{journal}{Nature} \textbf{\bibinfo{volume}{409}},
  \bibinfo{pages}{113} (\bibinfo{year}{2001}).

\bibitem[{\citenamefont{Skinner et~al.}(2004)\citenamefont{Skinner, Baumann,
  Quinn, Molloy, and Hoggett}}]{skinner2004promoter}
\bibinfo{author}{\bibfnamefont{G.~M.} \bibnamefont{Skinner}},
  \bibinfo{author}{\bibfnamefont{C.~G.} \bibnamefont{Baumann}},
  \bibinfo{author}{\bibfnamefont{D.~M.} \bibnamefont{Quinn}},
  \bibinfo{author}{\bibfnamefont{J.~E.} \bibnamefont{Molloy}},
  \bibnamefont{and} \bibinfo{author}{\bibfnamefont{J.~G.}
  \bibnamefont{Hoggett}}, \bibinfo{journal}{Journal of Biological Chemistry}
  \textbf{\bibinfo{volume}{279}}, \bibinfo{pages}{3239} (\bibinfo{year}{2004}).

\bibitem[{\citenamefont{Andrecka et~al.}(2008)\citenamefont{Andrecka, Lewis,
  Br{\"u}ckner, Lehmann, Cramer, and Michaelis}}]{andrecka2008single}
\bibinfo{author}{\bibfnamefont{J.}~\bibnamefont{Andrecka}},
  \bibinfo{author}{\bibfnamefont{R.}~\bibnamefont{Lewis}},
  \bibinfo{author}{\bibfnamefont{F.}~\bibnamefont{Br{\"u}ckner}},
  \bibinfo{author}{\bibfnamefont{E.}~\bibnamefont{Lehmann}},
  \bibinfo{author}{\bibfnamefont{P.}~\bibnamefont{Cramer}}, \bibnamefont{and}
  \bibinfo{author}{\bibfnamefont{J.}~\bibnamefont{Michaelis}},
  \bibinfo{journal}{Proceedings of the National Academy of Sciences}
  \textbf{\bibinfo{volume}{105}}, \bibinfo{pages}{135} (\bibinfo{year}{2008}).

\bibitem[{\citenamefont{Adelman et~al.}(2002)\citenamefont{Adelman, La~Porta,
  Santangelo, Lis, Roberts, and Wang}}]{adelman2002single}
\bibinfo{author}{\bibfnamefont{K.}~\bibnamefont{Adelman}},
  \bibinfo{author}{\bibfnamefont{A.}~\bibnamefont{La~Porta}},
  \bibinfo{author}{\bibfnamefont{T.~J.} \bibnamefont{Santangelo}},
  \bibinfo{author}{\bibfnamefont{J.~T.} \bibnamefont{Lis}},
  \bibinfo{author}{\bibfnamefont{J.~W.} \bibnamefont{Roberts}},
  \bibnamefont{and} \bibinfo{author}{\bibfnamefont{M.~D.} \bibnamefont{Wang}},
  \bibinfo{journal}{Proceedings of the National Academy of Sciences}
  \textbf{\bibinfo{volume}{99}}, \bibinfo{pages}{13538} (\bibinfo{year}{2002}).

\bibitem[{\citenamefont{Kipnis}(1986)}]{kipnis1986central}
\bibinfo{author}{\bibfnamefont{C.}~\bibnamefont{Kipnis}}, \bibinfo{journal}{The
  Annals of Probability} pp. \bibinfo{pages}{397--408} (\bibinfo{year}{1986}).

\bibitem[{\citenamefont{Ferrari and Fontes}(1996)}]{ferrari1996poissonian}
\bibinfo{author}{\bibfnamefont{P.}~\bibnamefont{Ferrari}} \bibnamefont{and}
  \bibinfo{author}{\bibfnamefont{L.}~\bibnamefont{Fontes}},
  \bibinfo{journal}{Journal of applied probability} pp.
  \bibinfo{pages}{411--419} (\bibinfo{year}{1996}).

\bibitem[{\citenamefont{Imamura and Sasamoto}(2007)}]{imamura2007dynamics}
\bibinfo{author}{\bibfnamefont{T.}~\bibnamefont{Imamura}} \bibnamefont{and}
  \bibinfo{author}{\bibfnamefont{T.}~\bibnamefont{Sasamoto}},
  \bibinfo{journal}{Journal of Statistical Physics}
  \textbf{\bibinfo{volume}{128}}, \bibinfo{pages}{799} (\bibinfo{year}{2007}).

\bibitem[{\citenamefont{Kolomeisky et~al.}(1998)\citenamefont{Kolomeisky,
  Sch{\"u}tz, Kolomeisky, and Straley}}]{kolomeisky1998phase}
\bibinfo{author}{\bibfnamefont{A.~B.} \bibnamefont{Kolomeisky}},
  \bibinfo{author}{\bibfnamefont{G.~M.} \bibnamefont{Sch{\"u}tz}},
  \bibinfo{author}{\bibfnamefont{E.~B.} \bibnamefont{Kolomeisky}},
  \bibnamefont{and} \bibinfo{author}{\bibfnamefont{J.~P.}
  \bibnamefont{Straley}}, \bibinfo{journal}{Journal of Physics A: Mathematical
  and General} \textbf{\bibinfo{volume}{31}}, \bibinfo{pages}{6911}
  (\bibinfo{year}{1998}).

\bibitem[{\citenamefont{Santen and Appert}(2002)}]{santen2002asymmetric}
\bibinfo{author}{\bibfnamefont{L.}~\bibnamefont{Santen}} \bibnamefont{and}
  \bibinfo{author}{\bibfnamefont{C.}~\bibnamefont{Appert}},
  \bibinfo{journal}{Journal of statistical physics}
  \textbf{\bibinfo{volume}{106}}, \bibinfo{pages}{187} (\bibinfo{year}{2002}).

\bibitem[{\citenamefont{R{\'a}kos et~al.}(2003)\citenamefont{R{\'a}kos,
  Paessens, and Sch{\"u}tz}}]{rakos2003hysteresis}
\bibinfo{author}{\bibfnamefont{A.}~\bibnamefont{R{\'a}kos}},
  \bibinfo{author}{\bibfnamefont{M.}~\bibnamefont{Paessens}}, \bibnamefont{and}
  \bibinfo{author}{\bibfnamefont{G.}~\bibnamefont{Sch{\"u}tz}},
  \bibinfo{journal}{Physical review letters} \textbf{\bibinfo{volume}{91}},
  \bibinfo{pages}{238302} (\bibinfo{year}{2003}).

\bibitem[{\citenamefont{Vuijk et~al.}(2015)\citenamefont{Vuijk, Rens, Vahabi,
  MacKintosh, and Sharma}}]{vuijk2015driven}
\bibinfo{author}{\bibfnamefont{H.~D.} \bibnamefont{Vuijk}},
  \bibinfo{author}{\bibfnamefont{R.}~\bibnamefont{Rens}},
  \bibinfo{author}{\bibfnamefont{M.}~\bibnamefont{Vahabi}},
  \bibinfo{author}{\bibfnamefont{F.~C.} \bibnamefont{MacKintosh}},
  \bibnamefont{and} \bibinfo{author}{\bibfnamefont{A.}~\bibnamefont{Sharma}},
  \bibinfo{journal}{Phys. Rev. E} \textbf{\bibinfo{volume}{91}},
  \bibinfo{pages}{032143} (\bibinfo{year}{2015}).

\bibitem[{\citenamefont{Ebbinghaus et~al.}(2011)\citenamefont{Ebbinghaus,
  Appert-Rolland, and Santen}}]{ebbinghaus2011particle}
\bibinfo{author}{\bibfnamefont{M.}~\bibnamefont{Ebbinghaus}},
  \bibinfo{author}{\bibfnamefont{C.}~\bibnamefont{Appert-Rolland}},
  \bibnamefont{and} \bibinfo{author}{\bibfnamefont{L.}~\bibnamefont{Santen}},
  \bibinfo{journal}{Journal of Statistical Mechanics: Theory and Experiment}
  \textbf{\bibinfo{volume}{2011}}, \bibinfo{pages}{P07004}
  (\bibinfo{year}{2011}).

\bibitem[{\citenamefont{Howard et~al.}(2001)}]{howard2001mechanics}
\bibinfo{author}{\bibfnamefont{J.}~\bibnamefont{Howard}} \bibnamefont{et~al.},
  \emph{\bibinfo{title}{Mechanics of motor proteins and the cytoskeleton}}
  (\bibinfo{publisher}{Sinauer Associates Sunderland, MA},
  \bibinfo{year}{2001}).

\bibitem[{\citenamefont{Spitzer}(1970)}]{spitzer1970interaction}
\bibinfo{author}{\bibfnamefont{F.}~\bibnamefont{Spitzer}},
  \bibinfo{journal}{Advances in Mathematics} \textbf{\bibinfo{volume}{5}},
  \bibinfo{pages}{246} (\bibinfo{year}{1970}).

\bibitem[{\citenamefont{Evans et~al.}(2003)\citenamefont{Evans, Juhasz, and
  Santen}}]{evans2003shock}
\bibinfo{author}{\bibfnamefont{M.}~\bibnamefont{Evans}},
  \bibinfo{author}{\bibfnamefont{R.}~\bibnamefont{Juhasz}}, \bibnamefont{and}
  \bibinfo{author}{\bibfnamefont{L.}~\bibnamefont{Santen}},
  \bibinfo{journal}{Physical Review E} \textbf{\bibinfo{volume}{68}},
  \bibinfo{pages}{026117} (\bibinfo{year}{2003}).

\bibitem[{\citenamefont{Cirillo et~al.}(2014)\citenamefont{Cirillo, Muntean,
  van Santen, and Sengar}}]{cirillo2014residence}
\bibinfo{author}{\bibfnamefont{E.~N.} \bibnamefont{Cirillo}},
  \bibinfo{author}{\bibfnamefont{A.}~\bibnamefont{Muntean}},
  \bibinfo{author}{\bibfnamefont{R.}~\bibnamefont{van Santen}},
  \bibnamefont{and} \bibinfo{author}{\bibfnamefont{A.}~\bibnamefont{Sengar}},
  \bibinfo{journal}{arXiv preprint arXiv:1411.5490}  (\bibinfo{year}{2014}).

\bibitem[{\citenamefont{Rajewsky et~al.}(1998)\citenamefont{Rajewsky, Santen,
  Schadschneider, and Schreckenberg}}]{rajewsky1998asymmetric}
\bibinfo{author}{\bibfnamefont{N.}~\bibnamefont{Rajewsky}},
  \bibinfo{author}{\bibfnamefont{L.}~\bibnamefont{Santen}},
  \bibinfo{author}{\bibfnamefont{A.}~\bibnamefont{Schadschneider}},
  \bibnamefont{and}
  \bibinfo{author}{\bibfnamefont{M.}~\bibnamefont{Schreckenberg}},
  \bibinfo{journal}{Journal of statistical physics}
  \textbf{\bibinfo{volume}{92}}, \bibinfo{pages}{151} (\bibinfo{year}{1998}).

\bibitem[{\citenamefont{Ferrari and Fontes}(1994)}]{ferrari1994shock}
\bibinfo{author}{\bibfnamefont{P.~A.} \bibnamefont{Ferrari}} \bibnamefont{and}
  \bibinfo{author}{\bibfnamefont{L.}~\bibnamefont{Fontes}},
  \bibinfo{journal}{Probability Theory and Related Fields}
  \textbf{\bibinfo{volume}{99}}, \bibinfo{pages}{305} (\bibinfo{year}{1994}).

\bibitem[{\citenamefont{Cividini et~al.}(2014)\citenamefont{Cividini, Hilhorst,
  and Appert-Rolland}}]{cividini2014exact}
\bibinfo{author}{\bibfnamefont{J.}~\bibnamefont{Cividini}},
  \bibinfo{author}{\bibfnamefont{H.}~\bibnamefont{Hilhorst}}, \bibnamefont{and}
  \bibinfo{author}{\bibfnamefont{C.}~\bibnamefont{Appert-Rolland}},
  \bibinfo{journal}{Journal of Physics A: Mathematical and Theoretical}
  \textbf{\bibinfo{volume}{47}}, \bibinfo{pages}{222001}
  (\bibinfo{year}{2014}).

\bibitem[{\citenamefont{Katz et~al.}(1987)\citenamefont{Katz, Lebowitz, and
  Spohn}}]{katz34nonequilibrium}
\bibinfo{author}{\bibfnamefont{S.}~\bibnamefont{Katz}},
  \bibinfo{author}{\bibfnamefont{J.}~\bibnamefont{Lebowitz}}, \bibnamefont{and}
  \bibinfo{author}{\bibfnamefont{H.}~\bibnamefont{Spohn}}, \bibinfo{journal}{J.
  Stat. Phys} \textbf{\bibinfo{volume}{34}}, \bibinfo{pages}{497}
  (\bibinfo{year}{1987}).

\bibitem[{\citenamefont{Popkov et~al.}(2003)\citenamefont{Popkov, R{\'a}kos,
  Willmann, Kolomeisky, and Sch{\"u}tz}}]{popkov2003localization}
\bibinfo{author}{\bibfnamefont{V.}~\bibnamefont{Popkov}},
  \bibinfo{author}{\bibfnamefont{A.}~\bibnamefont{R{\'a}kos}},
  \bibinfo{author}{\bibfnamefont{R.~D.} \bibnamefont{Willmann}},
  \bibinfo{author}{\bibfnamefont{A.~B.} \bibnamefont{Kolomeisky}},
  \bibnamefont{and} \bibinfo{author}{\bibfnamefont{G.~M.}
  \bibnamefont{Sch{\"u}tz}}, \bibinfo{journal}{Physical Review E}
  \textbf{\bibinfo{volume}{67}}, \bibinfo{pages}{066117}
  (\bibinfo{year}{2003}).

\bibitem[{\citenamefont{Wang et~al.}(2014)\citenamefont{Wang, Pfeuty, Thommen,
  Romano, and Lefranc}}]{wang2014minimal}
\bibinfo{author}{\bibfnamefont{J.}~\bibnamefont{Wang}},
  \bibinfo{author}{\bibfnamefont{B.}~\bibnamefont{Pfeuty}},
  \bibinfo{author}{\bibfnamefont{Q.}~\bibnamefont{Thommen}},
  \bibinfo{author}{\bibfnamefont{M.~C.} \bibnamefont{Romano}},
  \bibnamefont{and} \bibinfo{author}{\bibfnamefont{M.}~\bibnamefont{Lefranc}},
  \bibinfo{journal}{Physical Review E} \textbf{\bibinfo{volume}{90}},
  \bibinfo{pages}{050701} (\bibinfo{year}{2014}).

\bibitem[{\citenamefont{Neri et~al.}(2011)\citenamefont{Neri, Kern, and
  Parmeggiani}}]{Neri2011totally}
\bibinfo{author}{\bibfnamefont{I.}~\bibnamefont{Neri}},
  \bibinfo{author}{\bibfnamefont{N.}~\bibnamefont{Kern}}, \bibnamefont{and}
  \bibinfo{author}{\bibfnamefont{A.}~\bibnamefont{Parmeggiani}},
  \bibinfo{journal}{Phys. Rev. Lett.} \textbf{\bibinfo{volume}{107}},
  \bibinfo{pages}{068702} (\bibinfo{year}{2011}).

\bibitem[{\citenamefont{Neri et~al.}(2013)\citenamefont{Neri, Kern, and
  Parmeggiani}}]{Neri2013exclusion}
\bibinfo{author}{\bibfnamefont{I.}~\bibnamefont{Neri}},
  \bibinfo{author}{\bibfnamefont{N.}~\bibnamefont{Kern}}, \bibnamefont{and}
  \bibinfo{author}{\bibfnamefont{A.}~\bibnamefont{Parmeggiani}},
  \bibinfo{journal}{New Journal of Physics} \textbf{\bibinfo{volume}{15}},
  \bibinfo{pages}{085005} (\bibinfo{year}{2013}).

\end{thebibliography}

\end{document}